\documentclass[natbib=false,review=false]{jfp-epi}

\usepackage{sty/style}
\usepackage{sty/thesis}

\jfpVolume{xx}
\jfpArticle{x}
\jfpDOI{12345.67890}
\jfpYear{2026}
\jfpMonth{1}

\received[Submitted]{2026-xx-xx}
\received[accepted]{2xxx-xx-xx}

\title{Enhancing a Hierarchical Graph Rewriting Language based on MELL Cut Elimination}

\begin{document}

\author{Kento Takyu}
\orcid{0009-0009-4785-2378}
\takyurevise{
\affiliation{%
  \institution{Waseda University}
  \city{Tokyo}
  \country{Japan}
  \authoremail{k.takyu@ntt.com}
}
\authornote{\takyurevise{Current affiliation: Social Informatics Laboratories, NTT, Tokyo, Japan.}}
}
\author{Kazunori Ueda}
\orcid{0000-0002-3424-1844}  
\affiliation{%
  \institution{Waseda University}
  \city{Tokyo}
  \country{Japan}
  \authoremail{ueda@waseda.jp}
}

\begin{abstract}
Hierarchical graph rewriting is a highly expressive 
computational formalism that manipulates graphs 
enhanced with
box structures for representing hierarchies.
It has provided the foundations of various graph-based modeling tools,
but the design of 
high-level declarative languages based on
hierarchical graph rewriting is still a challenge.
For a solid design choice, well-established 
formalisms with backgrounds
other than
graph rewriting would provide useful guidelines.
Proof nets of Multiplicative Exponential Linear Logic (MELL) is such a
framework because its original formulation of cut elimination is 
essentially graph rewriting involving 
box structures, 
where \takyurevise{the} 
so-called promotion boxes 
with an indefinite number of non-local edges may be cloned, migrated and deleted.
This work builds on LMNtal
\ueda{(pronounced ``elemental'')}
 as a declarative language based on
hierarchical (port) graph rewriting, and discusses how it can be
extended to support the above operations on promotion boxes
of MELL proof nets.  LMNtal thus extended turns out to be a practical
graph rewriting language that has 
\takyurevise{a} 
strong affinity with MELL proof nets.  
The language features provided are general enough to
encode other well-established models of concurrency.
Using
the toolchain of LMNtal that provides state-space search and model checking,
we implemented cut
elimination rules of MELL proof nets in extended LMNtal and
demonstrated that the platform could serve as a useful workbench for
proof nets.
\end{abstract}

\maketitle

\section{Introduction}\label{section: intro}

Connectivity and hierarchy are the two major structuring mechanisms
that often occur simultaneously
in modeling various phenomena ranging from computing to human societies.
In organizational charts and system configuration
diagrams, connectivity is represented by lines, whereas
the hierarchy of groups is represented by box-like structures.
Because of their universality,
it is desirable to have a computational model and a programming language that
formally handles this structure which we call \textit{hierarchical graphs}. 

There exist diverse 
\textit{formalisms} for hierarchical graph rewriting, e.g., by
\cite{berry_chemical_1992}, 
\cite{milner_bigraphical_2001},
\cite{ueda_lmntal_2009},
\cite{drewes_hierarchical_2002},
\cite{ene_attributed_2018},
\cite{alves_new_2011}, and
\cite{muroya_hypernet_2020}.
\takyurevise{(These formalisms are surveyed in \cref{section: lmntal} and discussed as related work in \cref{section: related}.)} 
However, designing a practical high-level declarative 
\textit{language} based
on hierarchical graph rewriting is still a challenge.
Well-established formal systems potentially related to graph rewriting
could provide guidelines 
\takyurevise{for} 
a solid design. 
Multiplicative Exponential Linear Logic (MELL)
by \cite{girard_linear_1987} is such a 
formalism
related to hierarchical graph rewriting, in which
\textit{proof nets} were
formulated 
for
representing proofs graphically, abstracting some of the symmetries
in sequent-based proofs.
In particular, cut elimination rules for
exponentials are represented
as the migration, cloning, and deletion of
box structures called \textit{promotion boxes}.

In this study, we start with the hierarchical (port) graph rewriting
language LMNtal \cite{ueda_lmntal_2009} and refine
its language constructs to establish 
\takyurevise{a} 
strong affinity with the 
operations on proof nets.
We build on LMNtal 
because, 
unlike many other formalisms and languages 
\takyurevise{based} 
on graph rewriting,
it is formalized in the standard style of
programming languages with abstract syntax and structural operational
semantics. 
It also allows interpretation as a logic programming language
based on intuitionistic linear logic \cite{ueda_lmntal_2009}.

An encoding of MELL proof nets using the original constructs of LMNtal
was outlined in a poster \cite{takyu_encoding_2023},
showing a promising direction,
but 
we show that 
the operations on proof nets
can be fully supported only with newly designed and implemented constructs.
Formalization of graph cloning and deletion 
\takyurevise{has also been a challenge} 
%
in the \takyurevise{algebraic approach to graph rewriting} 
(\cite{corradini_sesqui-pushout_2006,overbeek_graph_2021,brenas_verifying_2018}\takyurevise{; see \cref{section: related} for further discussion} 
), but
the main theme of 
the present
work is to introduce the constructs for graph
cloning and deletion through boxes into a practical declarative language.

The main contributions of this paper are threefold.
Firstly, we extend 
the hierarchical graph rewriting
language LMNtal to make it possible to describe the operations of
promotion boxes in MELL proof nets, define operations corresponding to
the duplication and deletion of promotion boxes, and implement them 
(\cref{section: ext}). 
Secondly, we demonstrate that the extended LMNtal serves as a
workbench for proof nets (\cref{section: encoding-pn}).
Thirdly, we demonstrate that the proposed constructs are sufficiently
general by showing an encoding of the Ambient Calculus, a model of
concurrency with box-like structures
(\cref{section: encoding-proc}). 

%
\takyurevise{
This is a revised, extended version of the conference paper by the
authors \cite{takyu_PADL}.  The two principal additions in the
present article are as follows.
\begin{itemize}
\item \cref{subsec: push-eq} is a new subsection that demonstrates
  the use of our extended LMNtal as a workbench for proof nets,
  in which rewrite rules and equivalences can be easily modified or
  added, with their consequences observed through state-space
  exploration.  As a case study, we examine push/pull rules for
  weakening and explore candidate sufficient conditions for their
  safe use.
\item \cref{section: encoding-proc} gives, in a self-contained
  manner, the full LMNtal encoding of the Ambient Calculus
  (\cref{code:ambient}), with a focus on the cloning of replicated
  processes via $!(\texttt{open}\;m.P)$.  This shows that the
  proposed \texttt{mell} library is useful beyond proof nets.
\end{itemize}
In addition, \cref{section: lmntal} (LMNtal) and \cref{section: mell}
(MELL proof nets) have also been substantially expanded for
self-containedness, and \cref{section: ext} (the \texttt{mell}
library) spells out the motivation through a detailed comparison
with the pre-existing \texttt{nlmem} library.  The correspondence
between the LMNtal encodings and the proof nets/ambient calculus is
presented informally (through figures and rule-by-rule mapping); a
formal proof of the correspondence is beyond the scope of the
present paper.}

\section{LMNtal: A Hierarchical Graph Rewriting Language}\label{section: lmntal}


explain the hierarchical graph rewriting
language LMNtal 
\cite{ueda_lmntal_2009},
which consists of (i) term-based syntax,
(ii) structural congruence on terms that provides 
\takyurevise{an} 
interpretation of
terms as graphs, and (iii) small-step reduction relation.
A tutorial introduction to LMNtal can be found in \cite{LMNtal_tutorial}
and
the full description with formal definition in \cite{ueda_lmntal_2009}.

Let us start with 
\takyurevise{a} 
historical account of the language and its relation
to other formalisms and languages.
LMNtal is a computational model and a programming/modeling language
that uses logical 
(i.e., immutable) variables to represent connectivity and membranes to
represent hierarchy.  It is an outcome of the attempt to unify
constraint-based concurrency (a.k.a. concurrent constraint
programming) \cite{ueda-TACS2001}  
and Constraint Handling Rules (CHR) \cite{CHR}, 
two notable extensions 
\takyurevise{of} 
logic programming.
As graphs are a highly general structure, LMNtal is necessarily
related to many other computational models and languages.  The
following list would help 
the positioning of LMNtal
within a diverse spectrum of related formalisms.

\textbf{Logic Programming.} 
  Standard logic programming handles both
  function and predicate symbols, following first-order logic.
  The conception of LMNtal started with unifying them by representing
  $n$-ary function symbols in a relational form (with $n+1$ arguments).
  This resulted in doing away with the notion of ``instantiating''
  variables to non-variable terms.
  The role of variables in LMNtal is degenerated to indicating
  connections between atomic formulas.

\textbf{Constraint Handling Rules.}
  The above-mentioned adoption of a relational form results in
  rewrite rules with non-clausal form that allows
  multiple atomic formulas in the
  LHS, which can be found also in CHR.
  Both LMNtal and CHR could be thought of as
  an extension to multiset rewriting, but 
  CHR
  retains function symbols while LMNtal does not.
  Another difference is that the
  applications of LMNtal have been expanded towards computer-aided
  verification that involves state-space search and model checking.

\textbf{Linear Logic.}
  LMNtal was born from logic programming with
  don't-care nondeterminism 
  adopted by concurrent logic languages and CHR---%
  as opposed to don't-know nondeterminism (involving search)
  adopted by other logic programming languages.
  Don't-care nondeterminism (a.k.a. choice nondeterminism) is
  an important construct for modeling concurrency, but
  its logical interpretation had always been the point of interest.
  Following the linear-logic interpretation of CHR \cite{CHR-LL},
  \cite{ueda_lmntal_2009} provides the logical interpretation of
  don't-care nondeterminism based on the additive construct of 
  linear logic.

\textbf{Term Rewriting Systems.}
  In principle, graph rewriting naturally subsumes term rewriting, and
  a notable feature of LMNtal is that the rewriting is based on
  subgraph matching modulo structural congruence.  One difference
  is that, developed as a model of concurrency, LMNtal programs
  in general do not care about confluence or termination.

\textbf{Interaction Nets}
  \cite{lafont_interaction_1989}  
  could be considered as a minimal formalism that
  greatly simplifies concurrent graph rewriting
  by introducing the notion of
  principal ports to ensure confluence.
%
\takyurevise{
LMNtal can be regarded as a formalism that lifts the syntactic
conditions of Interaction Nets
\ueda{(i.e., binary interaction of nodes connected via principal ports)
by allowing LHS subgraphs which may be non-binary, need not be
connected, and allow hierarchical graphs.}}

\takyurevise{
\textbf{Port Graphs and PORGY.}
  Port graph rewriting and the PORGY system
  \cite{pinaud_porgy_2012} share \ueda{%
  LMNtal's motivation for adopting a port-graph-based approach}.
  LMNtal differs in (i) treating membranes as first-class
  hierarchies 
  (cf. Attributed Hierarchical Port Graphs 
  \ueda{%
  \cite{ene_attributed_2018}
  that have a quite different formulation)}
  and (ii) being formalized as a
  programming language with structural operational semantics;
  see \cref{section: related} for further discussion.
}

\textbf{Functional Programming and the $\lambda$-Calculus.}
  $\lambda$-terms can be represented naturally as graphs by using two
  kinds of edges, one for forming term structures and the other for
  forming binding structures.  This idea led us to develop several
  encodings of the $\lambda$-calculus with strong reduction, ranging
  from fine-grained ones (e.g., by \cite{ueda_encoding_2008})
  to coarse-grained ones (e.g., by \cite{Alim-Access}).
  Thus we can say that, although LMNtal does not feature built-in
  higher-order functions, they can be explicitly encoded using graphs and 
  a set of rewrite rules for $\beta$-reduction.

%
\takyurevise{
\textbf{Algebraic Graph Transformation.}
Among the formal approaches to graph transformation, there is a
well-established algebraic line of research \cite{EhrigEPT06},
where a rewrite step is expressed as a span of morphisms in a graph
category.  
The Double-Pushout (DPO) approach is a widely-known
formalism in this tradition; its variants such as Single-Pushout
(SPO), Sesqui-Pushout (SqPO) \cite{corradini_sesqui-pushout_2006},
and PBPO+ \cite{overbeek_graph_2021} extend DPO to support graph
cloning.  
LMNtal was developed quite independently of this
categorical tradition; its term-based formulation, with structural
congruence inducing graphs, offers a complementary view.  
}

We conclude this overview with mentioning other recent developments,
which includes (i) HyperLMNtal \cite{hyperlmntal}, \cite{Alim-Access},
a hypergraph extension to LMNtal;
(ii) QLMNtal \cite{mishina_introducing_2024}
that introduces quantification into rewriting and graph matching; 
and (iii) $\lambda_{GT}$ \cite{sano-icgt2023}, a purely functional language
with HyperLMNtal-like graphs as first-class data.


\subsection{Syntax of LMNtal}\label{section: lmntal_overview}

We describe and motivate the syntactic constructs of LMNtal.

The syntax of LMNtal is given in \cref{fig:lmntal-syntax}, where 
three
syntactic categories, \textit{link names} (denoted by $X$), 
\textit{atom names} (denoted by $p$), 
and possibly empty \textit{membrane names} (denoted by $m$),
are presupposed. 

\begin{figure}[t]
  \[
\begin{array}{r@{~~}c@{~~}r@{~~}l@{~~}l@{~~}l@{~~}l@{~~}l@{~~}l@{~~}l@{~~}l@{~~}l} \hline\\[-9pt]
    (\textit{process}) & P &::=& \zero &\bigm|& p\paren{X_1 \pc \ldots \pc X_n}&\bigm|& P \pc P &\bigm|& m\mem{P} &\bigm|& T \react T \\[3pt]
    (\textit{process template}) & T &::=& \zero &\bigm|& p\paren{X_1 \pc \ldots \pc X_n}&\bigm|& T \pc T &\bigm|& m\mem{T} &\bigm|& T \react T \\[1pt]
     && \bigm| & \verb+@+ p &\bigm|&\multicolumn{7}{l}{\!\!\!\texttt{\$} p\bracket{X_1 \pc \ldots \pc X_n \texttt{|} A}~~\bigm|~~p \paren{\texttt{*}X_1 \pc \ldots \pc \texttt{*}X_n}} \\[3pt]
    (\textit{residual}) & A &::=& \verb+[]+ &\bigm|&\texttt{*} X &&&&&&\\[3pt]
 \hline
  \end{array}
  \]%
   \vspace{-.5em}
   \caption{Syntax of LMNtal.}
  \label{fig:lmntal-syntax}
\end{figure}

Since LMNtal was originally developed as a model of concurrency,
the hierarchical graphs of LMNtal are also called \textit{processes}.
$\zero$ is an inert process; 
$p\paren{X_1 \pc \ldots \pc X_n} (n \geq 0)$ is an $n$-ary
\textit{atom} (a.k.a. node)
with \textit{ordered links} (a.k.a. edges) $X_1, \ldots, X_n$;
$P\pc P$ is parallel composition;
$m\mem{P}$ 
is a \textit{cell} formed by wrapping $P$ with an
optionally named \textit{membrane} 
\verb+{+~\verb+}+; and
$T \react T$ is a \textit{rewrite rule}.

Occurrences of a link name represent endpoints of a one-to-one link
between atoms (or more precisely, atom arguments).
%
\takyurevise{
  For this purpose, the following Link Condition is imposed on processes and rules:
\begin{itemize}
  \item Each link name 
\ueda{in a process $P$, excluding those appearing in rewrite rules, may}
occur at most twice in 
\ueda{$P$},
and each link name in a \ueda{rewrite} 
rule must occur exactly twice in the rule.
\end{itemize}
}
A link whose name
occurs only once in $P$ is called a \textit{free link} of $P$,
and a link whose name occurs exactly twice in $P$ is
called a \textit{local link} of $P$.
Links
may cross 
membranes and connect atoms located at different ``places'' of the
membrane hierarchy. 
A graph in which each node has its own arity and
totally ordered links,
like an LMNtal graph, is often called a \textit{port graph}. 
We adopt port graphs rather than ordinary graphs because the former
has better affinity with related notions (e.g., terms and formulas)
of programming languages and logic.  Still, nodes of ordinary graphs
with an indefinite number of unordered edges can be represented using membranes
with an indefinite number of free links.

\begin{example}
\item
A process
\begin{lstlisting}
  a(L1,F), b(L1,L2,L3,L4), c(L2,L5,L6,L6), d(L5,L3,L4)
\end{lstlisting}
stands for the undirected graph shown in \cref{fig:lmntalexample},
or more precisely, the \textit{undirected port
multigraph}, i.e., a graph that allows multi-edges and self-loops.
When a process is interpreted as a graph, 
the associativity of `\texttt{,}' for parallel composition is
insignificant (hence parentheses have been omitted) as well as
the ordering of the atoms (\texttt{a}--\texttt{d}).
Also, the names of local links (\texttt{L2}--\texttt{L6}) 
are insignificant.
These properties are represented as Structural Congruence\takyurevise{,} 
defined and discussed in \cref{sec:SC_LMNtal}.
\end{example}
\begin{figure}[t]
  \vspace{-1em}
  \centering
  \scalebox{0.9}{\input{figures/appendix/A/lmntalexample.tex}}
  \caption{Pictorial representation of an LMNtal graph,
  in which $\texttt{F}$ is a free link and the others are local
  links.
  An arrowhead of each non-unary atom
  indicates the first argument and the ordering 
  (clockwise or anticlockwise) of atoms.}
  \label{fig:lmntalexample}
\end{figure}

Process templates on both sides of a rewrite rule allow 
\textit{process contexts}, \textit{rule contexts}, and
\textit{aggregates} 
\cite{ueda_lmntal_2005,ueda_lmntal_2009}.

A process context, denoted $\verb+$+p\texttt{[}X_1 \pc \ldots \pc 
X_n \texttt{|} A\texttt{]}\ (n \geq 0) $, works as a 
\textit{wildcard} that
matches ``the rest of the
processes'' (except rewrite rules
that are matched by a rule context $\texttt{@}p$)
within the membrane in which it
appears. The arguments specify what free link names may or must
occur. $X_1 \pc \ldots \pc X_n$ are the link names that must occur
free in 
(the process that matches)
$\procvar p$.  
When the 
\textit{residual}
$A$ is of the form $\verb+*+X$ (\textit{bundle}),
links other than $X_1 \pc \ldots \pc X_n$ may occur free, and
$\verb+*+X$ stands for the 
\textit{(possibly empty)}
sequence of those optional free
links. When $A$ is of the form \verb+[+\verb+]+
(in which case the ``\texttt{|[]}'' may be omitted), 
no other free links
may occur. 

An aggregate, denoted 
$p\paren{\texttt{*}X_1 \pc \ldots \pc \texttt{*}X_n}\
(n > 0)$, represents a multiset of atoms with the name $p$, whose
multiplicity coincides with the number of links represented by the
argument bundles, denoted $|\texttt{*}X_i|$,
which must be uniform for $1\le i \le n$.
%

Rewrite rules must observe several syntactic conditions 
on links, (rule and process) 
contexts, aggregates, and bundles
\cite{ueda_lmntal_2009} so that
\begin{enumerate}
\item the Link Condition 
%
is preserved in the course of program execution and
\item matching of process contexts against processes is uniquely determined.
\end{enumerate}
Most importantly, in order to satisfy the former,
link names in each rewrite
rule must occur exactly twice, and 
to satisfy the latter,
each process context must occur
exactly once at the top level of distinct cells in the LHS of a rule.
It is worth noting that a rewrite rule containing 
rule contexts, process contexts and/or aggregates can be thought of 
as \textit{rule schemes} representing a set of rules 
obtained by instantiating 
those  
\takyurevise{wildcard} 
\takyurevise{constructs to concrete processes} 
they represent.


A rewrite rule repeatedly acts on an LMNtal graph located in the same
``place'' of the membrane hierarchy. 

\begin{example}\label{ex:first}
As an example of rewriting using process contexts and bundles,
consider the following program:
\begin{lstlisting}
  o(B), {i(A,B), a(A,G1), b(G2)}, g(G1), g(G2).  // initial graph
  o(B), {i(A,B), $p[A|*X]} :- n(A), $p[A|*X].    // rewrite rule
\end{lstlisting}
The first line is a text representation of the 
graph 
\takyurevise{in} 
\cref{figure: lmngraph}(left).  
%
%
The rule in the second line 
(\cref{figure: lmngraph}, center) rewrites this to the graph of 
\cref{figure: lmngraph}(right),
%
where the process context \verb+$p+ matches atoms 
$\texttt{a(A,G1)}$ and $\texttt{b(G2)}$,
and the bundle \verb+*X+ matches the free links $\texttt{G1}$, $\texttt{G2}$
of
\verb+$p+.
\end{example}

\begin{figure}[t]
\hbox to\hsize{%
    \hfill
    \scalebox{0.44}{\input{figures/2/bundle-before.tex}}
    \hfill
    \raisebox{15pt}{%
\scalebox{0.44}{\input{figures/2/bundle-rule.tex}}}
    \hfill
    \scalebox{0.44}{\input{figures/2/bundle-after.tex}}
    \hfill}
    \caption{Rewriting an LMNtal graph using a process context and a bundle.}
    \label{figure: lmngraph}
\end{figure}


The application of rules in LMNtal is nondeterministic because (i) a
rule may be able to rewrite different subgraphs of a given graph, and
(ii) different rules may be able to rewrite the same graph.
%
%
The LMNtal runtime SLIM
provides a \textit{nondeterministic execution
mode} that constructs the whole state space
of rewriting, which
%
%
can then be visualized using the visualization tool StateViewer
\cite{_lmntal_2010} (see \cref{subsection: lavit}).
Furthermore, SLIM provides an LTL model checker of the state space
\cite{gocho_evolution_2011}.
An important feature of SLIM 
is that (i) the rewriting is based on
subgraph matching \textit{up to structural congruence}
and that (ii) the construction of a state
space is again up to structural congruence, which can be interpreted as
graph isomorphism formulated in the LMNtal framework.  
In other words, our tool
comes with a built-in symmetry reduction mechanism.

Here are some notes on the syntax of LMNtal.

\begin{enumerate}
\item
\textit{Parallel composition} $P_1,P_2$
glues two processes $P_1$ and $P_2$ to build a larger process.
Note that, if each of 
$P_1$ and $P_2$ has a free link with the same name, it becomes a local
link in $(P_1, P_2)$.
A reader may notice that the Link Condition may not
  always allow us to form $(P_1, P_2)$ from
  two graphs $P_1$ and $P_2$ each satisfying the Link Condition.
  How LMNtal handles this point will be discussed in
  \cref{sec:SC_LMNtal}. 

\item
For readability, parallel composition may be written in a
period-terminated form as well as in a comma-separated form.  For instance,
$\texttt{a, (a:-b,c)}$ may be written as $\texttt{a. a:-b,c.}\,$,
where the comma binds tighter than `\texttt{:-}' which in turn
binds tighter than periods
\ueda{(readers may notice that this convention was used
in \cref{ex:first})}.

\item
A special binary atom, called a \textit{connector} $\texttt{=}(X,Y)$,
also written as $X\equals Y$,
fuses (or glues) two links $X$ and $Y$.
This is used, for example, in the encoding of
Rule (\textit{ax}-\textit{cut}) in \cref{code: full}.

\item 
  As a practical extension to the original definition
  \cite{ueda_lmntal_2009}, the present syntax and our
  implementation allow
  named membranes of the form $m\mem{\,}$ (as in
  \cref{section: lmntal_overview}) and
  named rules of the form
  $\textit{rule name}\texttt{@@}\,T\react T$
  (as in \cref{section: encoding-pn}).

\item
  Although not detailed here, our implementation comes with a 
  \textit{module
  system} used by application programming interfaces (APIs).
  Accordingly, an atom name can be prefixed by a module name,
  like $\textit{modulename}.\textit{atomname}$.

\item
A term representing a process is subject to Structural Congruence\takyurevise{,} 
defined in \cref{sec:SC_LMNtal},
which then stands for an \textit{undirected port
multigraph}, i.e., a graph that allows multi-edges and self-loops.
For instance, a process
\begin{lstlisting}
  a(L1,F), b(L1,L2,L3,L4), c(L2,L5,L6,L6), d(L5,L3,L4)
\end{lstlisting}
stands for the undirected graph shown in \cref{fig:lmntalexample},
where the order of atoms and the associativity of `\texttt{,}' for parallel
composition 
are not significant because of the structural
congruence described in \cref{sec:SC_LMNtal}.
\end{enumerate}

\subsubsection{Nonlinear Membranes}\label{subsection: nlmem}

When process contexts with a bundle in the LHS of a rule are cloned or
deleted in the RHS, 
each of the endpoints of the links matched by the bundle
must be connected to some atom to avoid dangling links.  Aggregates
are the construct provided for this purpose, but instead of
providing them in full generality, the LMNtal implementation has provided
its functionalities as a wrapper library called $\texttt{nlmem}$
(non-linear membrane) \cite{_lmntal_2008}.
%
%
\takyurevise{
The library implements $\texttt{nlmem.copy}$ (\cref{figure: nlmem}(a)),
\ueda{%
where 
$\texttt{nlmem.copy(A,N,B)}$ with $\texttt{n(N)}$ 
(\cref{figure: nlmem}(a)(left))
clones
the cell $\texttt{\{\$p[A|*X]\}}$
into two, with fresh $\texttt{n}$ atoms
inserted between corresponding free links of the two clones
(\cref{figure: nlmem}(a)(right)).
Here, the first argument 
$\texttt{A}$ is a link to the 
cell to be cloned, i.e.,
the named link of $\texttt{\$p}$; 
the second argument 
$\texttt{N}$ is a link to an
auxiliary atom $\texttt{n(N)}$ whose name, here $\texttt{n}$, is used as
the wrapper atom name in the result; and 
the third argument 
$\texttt{B}$ is the access
link to the two clones after the rewrite.}

For example, when \texttt{\$p} stands for $\texttt{a(A,X1,X2)}$ (i.e.,
$|\texttt{*X}| = 2$), the rewrite} 
\begin{lstlisting}[basicstyle=\ttfamily\small]
  nlmem.copy(A,N,B), n(N), {a(A,X1,X2)}.
\end{lstlisting}
\takyurevise{produces} 
\begin{lstlisting}[basicstyle=\ttfamily\small]
  {a(A1,X3,X4)}, {a(A2,X5,X6)}, n(X1,X3,X5), n(X2,X4,X6), n(A1,A2,B).
\end{lstlisting}
\takyurevise{The fresh $\texttt{n}$ atoms inserted between corresponding free links
of the two clones share the atom name with the auxiliary
$\texttt{n(N)}$ in the LHS.} 
Similarly, the API provides
$\texttt{nlmem.kill}$ (\cref{figure: nlmem}(b)), which
removes a membrane with an unspecified number of free links.
%
Note that the semantics of these operations can be defined using
aggregates \cite{_lmntal_2005}.

\begin{figure}[t]
    \newlength{\cheight}
    \settoheight{\cheight}{\scalebox{0.488}{\input{figures/2/nlmem2.tex}}}
  \begin{minipage}[t]{0.53\hsize}
    \begin{center}
      \scalebox{0.488}{\input{figures/2/nlmem2.tex}}\\[3pt]
      \texttt{\small nlmem.copy(A,N,B),n(N),\{\$p[A|*X]\}}\\[3pt]
      (a)
    \end{center}
  \end{minipage}
  \hfill
  \begin{minipage}[t]{0.46\hsize}
    \begin{center}
      \raisebox{0.2\cheight}{\scalebox{0.488}{\input{figures/2/nlmem1.tex}}}\\[3pt]
      \texttt{\small nlmem.kill(A,N),n(N),\{\$p[A|*X]\}}\\[3pt]
      (b)
    \end{center}
  \end{minipage}
  \caption{Operations of the \texttt{nlmem} library.}
  \vspace{-0.75em}
  \label{figure: nlmem}
\end{figure}

\subsection{Semantics of LMNtal}\label{sec:semantics_LMNtal}

The semantics of LMNtal consists of 
\takyurevise{a} 
\textit{structural congruence} 
(\cref{sec:SC_LMNtal}) and a \textit{reduction relation}
(\cref{sec:reduction_relation}).

\subsubsection{Structural Congruence}\label{sec:SC_LMNtal}

The syntax defined 
in \cref{section: lmntal_overview}
does not yet characterize LMNtal graphs
because the port graph of \cref{fig:lmntalexample} allows
other syntactic representations.
\cref{fig:Structural congruence on LMNtal processes} 
defines an equivalence relation,
called
\textit{structural congruence}, to absorb the syntactic variations.
$P[Y/X]$ stands for the renaming of link $X$ in process $P$ to $Y$. 
The rules apply only when each process satisfies the Link Condition.
The rules are as in \cite{ueda_lmntal_2009} and readers familiar with
LMNtal may skip the details. 

\begin{figure}[t]
	\centering
	\begin{tabular}{rr@{~~}c@{~~}l@{~~}l}
		\hline\\[-9pt]
		(E1) & $\textbf{0},P$ & $\equiv$ & $P$ \\[1pt]
 		(E2) & $P,Q$ & $\equiv$ & $Q,P$ \\[1pt]
		(E3) & $P,(Q,R)$ & $\equiv$ & $(P,Q),R$ \\[1pt]
		(E4) & $P$ & $\equiv$ & $P[Y/X]$ & if $X$ is a local link of $P$ \\[1pt]
		(E5) & $P\equiv P'$ & $\Rightarrow$ & $P,Q\equiv P',Q$ \\[1pt]
		(E6) & $P\equiv P'$ & $\Rightarrow$ & $m\mem{P}\equiv m\mem{P'}$ \\[1pt]
		(E7) & $X\equals X$ & $\equiv$ & $\textbf{0}$ \\[1pt]
 		(E8) & $X\equals Y$ & $\equiv$ & $Y\texttt{=}X$ \\[1pt]
		(E9) & $X\equals Y,P$ & $\equiv$ & $P[Y/X]$ &
                       if $P$ is an atom and $X$ is a free link of $P$ \\[1pt]
		(E10) & $m\mem{X\equals Y,P}$ &
                $\equiv$ & $X\equals Y, m\mem{P}$ & 
                 if exactly one of $X$ and $Y$ is a free link of $P$ \\[3pt]
		\hline
	\end{tabular}
	\caption{Structural congruence on LMNtal processes.}
	\label{fig:Structural congruence on LMNtal processes}
\end{figure}

(E1)--(E3) characterize 
parallel composition of
atoms, 
membranes, and rules
as multisets.  
(E4) is  $\alpha$-conversion of local link names, where $Y$ must be a
fresh link because of the Link Condition.
When we compose two graphs with a comma, we regard the two graphs to
be $\alpha$-converted by (E4) as necessary to avoid collisions of
local link names\footnote{When each of $G_1$ and $G_2$ satisfies the
Link Condition but their composition
$G_1 \pc G_2$ does not, there must be a link occurring twice in one
and at least once in the other.  Since the former is
a local link, it can be renamed by (E4) to restore the Link Condition.}.
(E5) and (E6) are
structural rules to make $\equiv$ a congruence.
(E7)--(E10) are rules for connectors:
(E7) says that a self-closed link is regarded as an empty graph;
(E8) states the symmetry of connectors; 
(E9) says that a connector may be absorbed or emitted by an atom; and
(E10) says that connectors may move across membranes.


\subsubsection{Term Notation}

For convenience, the following term notation is allowed and is widely
used\footnote{%
This syntactic convention is explained here because the
example makes sense only after structural congruence has been introduced.};
that is, we allow
%
\[
p(X_1,\narrowdots,X_{k-1},L,X_{k+1},\narrowdots,X_m ),\,q(Y_1,\narrowdots,Y_{n-1},L)\quad
(1 \leq k \leq m,1 \leq n),
\]
to be written as
$p(X_1,\narrowdots,X_{k-1},q(Y_1,\narrowdots,Y_{n-1}),X_{k+1},\narrowdots,X_m),$
\ueda{where this embedding (of an atom into another atom)
may be done recursively.}

\begin{example}
$\texttt{p(X,Y,R),a(X),b(Y)}$ is a relational representation of a
term $\texttt{p(a,b)}$ rooted at $\texttt{R}$, 
which is made clear as follows:
It can be written as $\texttt{p(a,Y,R),b(Y)}$ and then as 
$\texttt{p(a,b,R)}$.
%
\takyurevise{
(E9) states that it is equivalent to 
$\texttt{R}\equals\texttt{Z,}\,\texttt{p(a,b,Z)}$.
By using the term notation again, with 
\ueda{$\texttt{R}\equals\texttt{Z}$} as the outer \ueda{(host)} atom
and $\texttt{p(a,b,Z)}$ as the
\ueda{embedded} atom, it can be written as $\texttt{R}\equals\texttt{p(a,b)}$,
which is more
concise and looks like a standard term.
} 
\end{example}

Similarly, we allow
\ueda{%
$
p(X_1,\narrowdots,X_{k-1},L,X_{k+1},\narrowdots,X_n ),\,
m\mem{\texttt{+}L,P}\> (1 \leq k \leq n)
$
to be written as
$
p(X_1,\narrowdots,X_{k-1},m\mem{P},X_{k+1},\narrowdots,X_n)\> (1 \leq k \leq n),
$}
\takyurevise{where $\texttt{+}L$ is a prefix-operator notation for a unary atom $\texttt{+}(L)$.} 

\subsubsection{Reduction Relation}\label{sec:reduction_relation}

\begin{figure}[t]
	\centering
	\begin{tabular}{l@{~}ll@{~}ll@{~}l}
		\hline\\[-2ex]
	(R1) & $\dfrac{P\reduces P'}{P,Q\reduces P',Q}\quad$ 
&	(R2) & $\dfrac{P\reduces P'}{m\{P\}\reduces m\{P'\}}\quad$
&	(R3) & $\dfrac{Q\equiv P\hspace{10pt}P\reduces P'\hspace{10pt}P'\equiv Q'}{Q\reduces Q'}$ \\[9pt]
%
(R4) & \multicolumn{5}{l}{$m\{X=Y,P\}\reduces X=Y,m\{P\}$\quad if $X$ and $Y$ are free links of \takyurevise{$\{ X{=}Y, P \}$}}\\[1pt]
(R5) & \multicolumn{5}{l}{$X=Y,m\{P\}\reduces m\{X=Y,P\}$\quad if $X$ and 
$Y$ are free links of $P$} \\[1pt]
(R6) & \multicolumn{3}{l}{$T\theta,(T\texttt{:-}U)\reduces U\theta,(T\texttt{:-}U)$} \\[3pt]
		\hline
	\end{tabular}
	\caption{Reduction relation on LMNtal processes.}
	\label{fig:Reduction relation on LMNtal processes}
\end{figure}

The reduction relation $\reduces$
is the minimum binary relation satisfying the rules
in \cref{fig:Reduction relation on LMNtal processes}.

(R1)--(R3) are standard structural rules:
(R1) and (R3) are indeed shared with other process calculi including
the $\pi$-calculus.  (R2) represents autonomous process evolution
inside a membrane.

%
(R4) and (R5) handle the pulling/pushing of connectors out of/into
membranes, respectively.
\ueda{Note that they decrease the number of links crossing
the membrane by two, unlike (E10) that 
lets a connector cross a membrane in zero steps when it 
does not change the number of crossing links.
The details of this design decision can be
found in \cite{ueda_lmntal_2009}, Section 5.2.}
The key rule, (R6), handles rewriting (of the process $T\theta$)
by a rule.  The substitution $\theta$,
determined upon graph matching,
is to map process contexts (wildcards for non-rule processes),
rule contexts (wildcards for rewrite rules), and aggregates
into specific processes, rules, and collections of atoms,
respectively.  The precise
(but somewhat tedious)
formulation of 
the substitution is left to \cite{ueda_lmntal_2009}.
\subsection{The LMNtal Toolchain}
\label{subsection: lavit}

The LMNtal toolchain comes with the following submodules:
\begin{enumerate}
  \item a compiler (written in Java) to the dedicated virtual machine code,
  \item SLIM, a virtual machine (written in C++) 
    supporting both ordinary execution and
    state space construction/search, the latter scaling up to $10^9$
    states,
  \item two graph visualizers (Unyo-Unyo and Graphene) for animated
    graph rewriting, and
  \item StateViewer that visualizes the state space of
    programs and counterexample paths of LTL model
    checking and allows the browsing of states and state transitions
    \cite{_lmntal_2010},
\end{enumerate}
all accessible from
LaViT\footnote{\url{https://www.uedalab.jp/lmntal/lavit/}}
(LMNtal Visual Tools),
an integrated development environment (IDE) of LMNtal.
LaViT allows one
to write, run, and model-check LMNtal programs with a simple user interface. 
In particular, StateViewer visualizes the state space of programs and
offers an environment for \textit{understanding} the properties of
programs with nondeterministic behavior
(unlike standard model checkers which just 
\takyurevise{return}  
``OK'' when no bugs
\takyurevise{are} 
found
).


\begin{example}\label{ex:lavit}
Consider the following LMNtal program:

\begin{lstlisting}
  a, b, b, b.
  a2c @@ a :- c.
  b2c @@ b :- c.
  bb2cc @@ b, b :- c, c.
\end{lstlisting}
\begin{figure}[t]
  \hfill
  \begin{minipage}[t]{0.32\hsize}
    \centering
    \scalebox{0.14}{\includegraphics{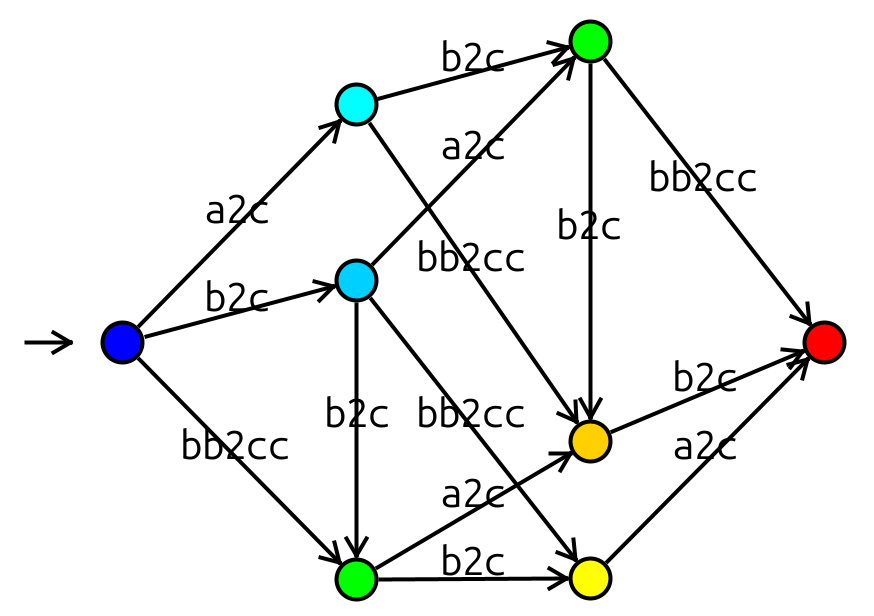}}
    \subcaption{Initial diagram}
  \end{minipage}
  \hfill
  \begin{minipage}[t]{0.32\hsize}
    \centering
    \scalebox{0.14}{\includegraphics{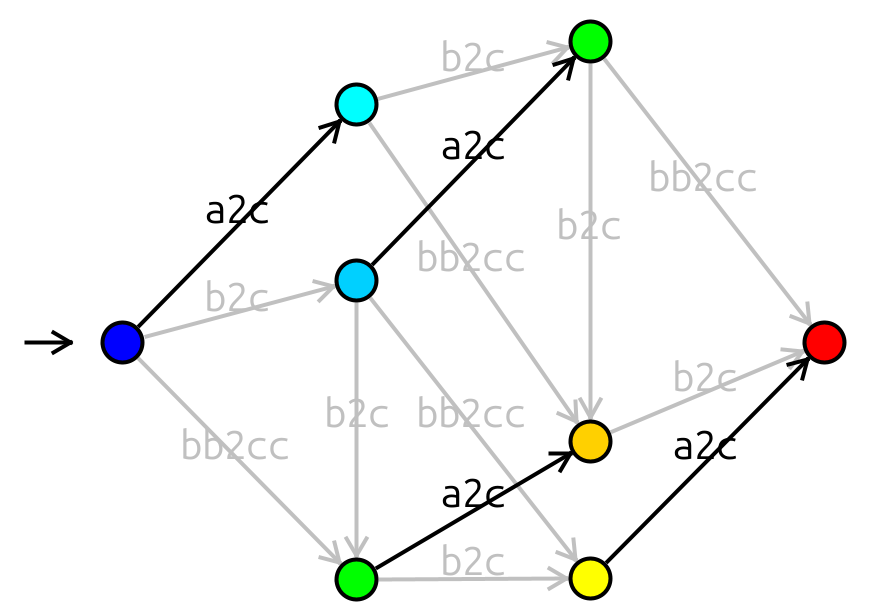}}
    \subcaption{Rule Name Search}
  \end{minipage}
  \hfill
  \begin{minipage}[t]{0.26\hsize}
    \centering
    \raisebox{10pt}{\scalebox{0.14}{\includegraphics{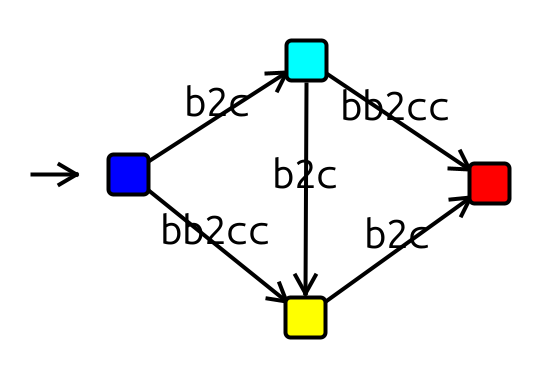}}}
    \subcaption{Transition Abstraction}
  \hfill
  \end{minipage}
  \caption{%
    State transition diagrams of an LMNtal program generated by LaViT:
(a) state space of the program of \cref{ex:lavit}; 
(b) state space with Rule Name Search of \texttt{a2c}; 
(c) state space with Transition Abstraction with \texttt{a2c}.%
  }
  \label{figure: stateviewer}
\end{figure}
%
With StateViewer, we can view a state space shown in \cref{figure: stateviewer}(a).
Each node represents an LMNtal graph.
StateViewer provides several features that help investigate the state space.
Clicking on each node displays the text representation of the
LMNtal graph.
%
One can also search for the
occurrence of a specific name in each state 
or the applied rule,
and the search results are highlighted on
the state space. 
\cref{figure: stateviewer}(b) highlights
the transitions
with the rule \texttt{a2c}
by Rule Name Search. 
Transition Abstraction 
visualizes state transitions with specified rules only, abstracting
transitions with other rules.
By choosing the rule
\texttt{a2c} and performing Transition Abstraction, we obtain a state
space shown in \cref{figure: stateviewer}(c),
in which small squares stand for abstracted states.
\end{example}

\section{MELL Proof Nets}\label{section: mell}

\subsection{Overview of MELL Proof Nets}

Proof nets were introduced by \cite{girard_linear_1987} in the context
of linear logic's proof theory, and have since been utilized in
various areas, including the analysis of the lambda calculus.

Proof nets are an abstraction of 
proof trees represented as graphs.
Proof trees may allow
essentially the same proofs to be
represented in
\takyurevise{a} 
different forms due to the variations in the
order of inference rule applications. 
Proof nets abstract away such syntactic differences and represent these proofs uniformly.
For example, in \cref{figure:pn-sample}, the left and right proof nets
use the same rules but in \takyurevise{a} 
different
order of application of the $\linpar$ and
$\otimes$ rules in the second and third final steps.
These are represented as the same proof net shown in the center. 

This study focuses on Multiplicative Exponential Linear Logic (MELL)
\cite{girard_linear_1987},
a fragment of linear logic which extends
Multiplicative Linear Logic (MLL) with
two exponential operators `$!$' and `$?$' 
to allow non-linear, ``classical'' handling of resources.

The exponential 
\takyurevise{operator `$!$' is} 
represented using a box structure called a
\textit{promotion box} in proof nets.
%
Box structures such as promotion boxes 
are sometimes a
subject of research towards 
dispensing with the
boxes
\cite{gonthier_linear_1992}, especially 
when rewriting of bounded-size redexes is pursued.
However, we consider them valuable to explore their relationship with
formalisms that explicitly or implicitly employ box structures in
their modeling frameworks,
including the $\lambda$-calculus whose standard formulation
  comes with substitution that may involve copying and deletion of
  unbounded-size redexes.

\begin{figure}[t]
  \begin{center}
       {
        \small
          \begin{minipage}{0.3\hsize}
            \scalebox{0.8}{
           $
           \begin{prooftree}[rule margin=0.5ex]
             \hypo{}
             \infer1[\textit{ax}]{\vdash n^{\bot}, n}
             \hypo{}
             \infer1[\textit{ax}]{\vdash n^{\bot}, n}
             \infer2[$\otimes$]{\vdash n^{\bot}, n^{\bot}, (n \otimes n)}
             \hypo{}
             \infer1[\textit{ax}]{\vdash n^{\bot}, n}
             \infer2[$\enf{\otimes}$]{\vdash n^{\bot}, n^{\bot}, n^{\bot}, (n \otimes n)\: \enf{\otimes}\: n}
             \infer1[$\alertt{\linpar}$]{\vdash (n^{\bot}\: \alertt{\linpar}\: n^{\bot}), n^{\bot}, (n \otimes n)\: \enf{\otimes}\: n}
             \infer1[$\linpar$]{\vdash (n^{\bot}\: \alertt{\linpar}\: n^{\bot}) \: \linpar \: n^{\bot}, (n \otimes n)\: \enf{\otimes}\: n}
           \end{prooftree}
           $
         }
        \end{minipage}
        \begin{minipage}{0.4\hsize}
          \begin{center}
            \begin{tikzpicture}
              \node (pn) at (0,0) {\scalebox{0.4}{\input{figures/3/pn-sample.tex}}};
              \node (el) at (-1.8,0) {\large$\mapsto$};
              \node (er) at (2,0) {\large$\lmapsto$};
              \end{tikzpicture}
          \end{center}
        \end{minipage}
        \begin{minipage}{0.28\hsize}
          \begin{flushleft}
            \hspace*{-1.2em}
            \scalebox{0.8}{
            $
            \begin{prooftree}[rule margin=0.5ex]
              \hypo{}
              \infer1[\textit{ax}]{\vdash n^{\bot}, n}
              \hypo{}
              \infer1[\textit{ax}]{\vdash n^{\bot}, n}
              \infer2[$\otimes$]{\vdash n^{\bot}, n^{\bot}, (n \otimes n)}
              \infer1[$\alertt{\linpar}$]{\vdash (n^{\bot}\: \alertt{\linpar}\: n^{\bot}), (n \otimes n)}
              \hypo{}
              \infer1[\textit{ax}]{\vdash n^{\bot}, n}
              \infer2[$\enf{\otimes}$]{\vdash (n^{\bot} \alertt{\linpar} n^{\bot}), n^{\bot}, (n \otimes n) \enf{\otimes} n}
              \infer1[$\linpar$]{\vdash (n^{\bot}\: \alertt{\linpar}\: n^{\bot}) \: \linpar \: n^{\bot}, (n \otimes n) \enf{\otimes} n}
\end{prooftree}
            $
            }
          \end{flushleft}
        \end{minipage}
      }
  \end{center}
  \caption{Two proof trees mapped to the same proof net.}\label{figure:pn-sample}
\end{figure}

\subsection{Multiplicative Exponential Linear Logic \takyurevise{with the MIX Rules}}
\label{subsection:mell}
\begin{definition}[MELL formula]
The MELL formulas $F$ 
\takyurevise{are} 
defined as follows:
\[
F ::= X\;\mid\;X^{\bot}\;\mid\;F\otimes F \;\mid\; F \:\linpar\: F \;\mid\; !F \;\mid\;?F
\]
where $X$ represents an atomic formula.
The negation $^{\bot}$ 
attached to non-atomic formulas
is moved/removed by:
\[
      A^{\bot \bot} := A \;\;\;\;\; (A\otimes B)^{\bot} := A^{\bot}\: \linpar\: B^{\bot} \;\;\;\;\;(A\:\linpar \: B)^{\bot} := A^{\bot} \otimes B^{\bot}
\]
\[
      (!A)^{\bot} :=\; ?(A^{\bot}) \;\;\;\;\;\;(?A)^{\bot} :=\; !(A^{\bot})
\]
\takyurevise{Linear} 
implication 
$\multimap$ is defined as follows:
\[
  A \multimap B := A^{\bot}\:\linpar\:B
\]
\takyurevise{Classical} 
implication $\to$ can be translated by using linear implication $\multimap$ and $!$:
\[
        A \to B \mapsto (!A \multimap B)  
\]
\end{definition}
  This paper does not consider multiplicative units.
  The handling of the units is discussed in \cite{accattoli_linear_2013}.
Figure~\ref{figure:LL-sequent} shows the inference rules of MELL
(one-sided).
Here, $A$ and $B$ are formulas. $\Gamma$ and $\Delta$ are
sequences of formulas by default,
but we implicitly allow exchange of formulas in $\Gamma, \Delta$ and
regard them as multisets.
%
Note the presence of the binary and nullary  
\textit{mix}
 rules \cite{fleury_retore_1994,accattoli_linear_2013}. 
  Allowing the mix rules relaxes the Danos-Regnier correctness (\cref{subsection: proofnets})
  and simplifies the definition of proof nets.

\begin{figure}[t]
	\hrulefill
        \begin{displaymath}
            \frac{\vdash \Gamma, A, B}{\vdash \Gamma, A\:\linpar\:B}(\linpar)\;\;\;\;\; \frac{\vdash \Gamma, A \;\;\;\;\; \vdash \Delta, B}{\vdash \Gamma, \Delta, A\otimes B}(\otimes)\;\;\;\;\;
            \frac{}{\vdash A, A^{\bot}}(\textit{ax})\;\;\;\;\;\frac{\vdash \Gamma, A \;\;\;\;\; \vdash \Delta, A^{\bot}}{\vdash \Gamma, \Delta}(cut)
        \end{displaymath}
        \vspace{-0.8em}
        \begin{displaymath}
            \frac{\vdash ?\Gamma, A}{\vdash ?\Gamma, !A}(!)\;\;\;\;\;\frac{\vdash \Gamma, A}{\vdash \Gamma, ?A}(?d)\;\;\;\;\;\frac{\vdash \Gamma}{\vdash \Gamma, ?A}(?w)\;\;\;\;\;\frac{\vdash \Gamma, ?A,?A}{\vdash \Gamma, ?A}(?c)
        \end{displaymath}
        \begin{displaymath}
          \frac{\;\;\;}{\vdash}(\textit{mix}_0)\;\;\;\;\;\ \frac{\vdash \Gamma \;\;\;\;\; \vdash \Delta }{\vdash \Gamma, \Delta}(\textit{mix}_2)
        \end{displaymath}
        \vspace{-0.5em}
	\hrulefill
        \vspace{0.5em}
  \caption{Sequent calculus for MELL (+ mix rules).}\label{figure:LL-sequent}
\end{figure}

It is known that the cut elimination theorem holds for the proof system of MELL \cite{girard_linear_1987}.
\cref{figure:LL-cut-full} shows the \takyurevise{principal} 
cut elimination rules of MELL.
In the inference rules, the application of rules shown with a double line means that the rule is applied as many times as necessary.
Although many rules do not immediately eliminate cuts, they help
elimination
by moving cuts towards the top of the proof tree. 

\begin{figure}[t]
      \centering
      \[
      \begin{prooftree}
        \infer0[(\textit{ax})]{ \vdash  A, A^{\bot}}
        \hypo{ \vdash \Gamma, A }
        \infer2[(\textit{cut})]{ \vdash \Gamma, A}
      \end{prooftree}
      \quad \rightsquigarrow \quad
      \begin{prooftree}
        \hypo{ \vdash \Gamma,  A }
      \end{prooftree}
      \]
      {\normalsize (\textit{ax}-\textit{cut})}
      \[
      \begin{prooftree}
        \hypo{ \vdash \Gamma , A, B } \infer1[($\linpar$)]{ \vdash \Gamma , A\: \linpar\: B }
        \hypo{ \vdash \Delta, A^{\bot} } 
        \hypo{ \vdash \Sigma, B^{\bot} } 
        \infer2[($\otimes$)]{ \vdash \Delta, \Sigma, A^{\bot} \otimes B^{\bot} }
        \infer2[(\textit{cut})]{ \vdash \Gamma, \Delta, \Sigma }
      \end{prooftree}
      \quad \rightsquigarrow \quad
      \begin{prooftree}
        \hypo{ \vdash \Gamma , A, B } 
        \hypo{ \vdash \Delta , A^{\bot} } 
        \infer2[(\textit{cut})]{ \vdash \Gamma , \Delta, B }
        \hypo{ \vdash \Sigma, B^{\bot} } 
        \infer2[(\textit{cut})]{ \vdash \Gamma, \Delta, \Sigma }
      \end{prooftree}
      \]
      {\normalsize ($\otimes \, \mathchar`- \, \linpar$)}
      \[
      \begin{prooftree}
        \hypo{ \vdash ?\Gamma , A } \infer1[($!$)]{ \vdash ?\Gamma , !A }
        \hypo{ \vdash ?\Delta, B, ?A^{\bot} } \infer1[($!$)]{ \vdash ?\Delta, !B, ?A^{\bot} }
        \infer2[(\textit{cut})]{ \vdash ?\Gamma, ?\Delta, !B }
      \end{prooftree}
      \quad \rightsquigarrow \quad
      \begin{prooftree}
        \hypo{ \vdash ?\Gamma , A } \infer1[($!$)]{ \vdash ?\Gamma , !A }
        \hypo{ \vdash ?\Delta, B , ?A^{\bot} }
        \infer2[(\textit{cut})]{ \vdash ?\Gamma, ?\Delta, B }
        \infer1[($!$)]{ \vdash ?\Gamma, ?\Delta, !B }
      \end{prooftree}
      \]
      {\normalsize ($! \, \mathchar`- \, !$)}
      \[
      \begin{prooftree}
        \hypo{ \vdash ?\Gamma , A } \infer1[($!$)]{ \vdash ?\Gamma , !A }
        \hypo{ \vdash \Delta, A^{\bot} } \infer1[($?d$)]{ \vdash \Delta, ?A^{\bot} }
        \infer2[(\textit{cut})]{ \vdash ?\Gamma, \Delta }
      \end{prooftree}
      \quad \rightsquigarrow \quad
      \begin{prooftree}
        \hypo{ \vdash ?\Gamma, A}
        \hypo{ \vdash \Delta, A^{\bot} }
        \infer2[(\textit{cut})]{ \vdash ?\Gamma, \Delta}
      \end{prooftree}
      \]
      {\normalsize ($! \, \mathchar`- \, ?d$)}
      \[
      \begin{prooftree}
        \hypo{ \vdash ?\Gamma , A } \infer1[($!$)]{ \vdash ?\Gamma , !A }
        \hypo{ \vdash \Delta } \infer1[($?w$)]{ \vdash \Delta, ?A^{\bot} }
        \infer2[(\textit{cut})]{ \vdash ?\Gamma, \Delta }
      \end{prooftree}
      \quad \rightsquigarrow \quad
      \begin{prooftree}
        \hypo{ \vdash \Delta}
        \infer[double]1[($?w$)]{ \vdash ?\Gamma, \Delta}
      \end{prooftree}
      \]
      {\normalsize ($! \, \mathchar`- \, ?w$)}
      \[
      \begin{prooftree}
        \hypo{ \vdash ?\Gamma , A } \infer1[($!$)]{ \vdash ?\Gamma , !A }
        \hypo{ \vdash \Delta, ?A^{\bot}, ?A^{\bot} } \infer1[($?c$)]{ \vdash \Delta, ?A^{\bot} }
        \infer2[(\textit{cut})]{ \vdash ?\Gamma, \Delta }
      \end{prooftree}
      \quad \rightsquigarrow \quad
      \begin{prooftree}
        \hypo{ \vdash ?\Gamma, A }
        \infer1[($!$)]{ \vdash ?\Gamma, !A }
        \hypo{ \vdash ?\Gamma, A }
        \infer1[($!$)]{ \vdash ?\Gamma, !A }
        \hypo{ \vdash \Delta, ?A^{\bot}, ?A^{\bot} }
        \infer2[(\textit{cut})]{ \vdash ?\Gamma, \Delta, ?A^{\bot} }
        \infer2[(\textit{cut})]{ \vdash ?\Gamma, ?\Gamma, \Delta}
        \infer[double]1[($?c$)]{ \vdash ?\Gamma, \Delta}
      \end{prooftree}
      \]
      {\normalsize ($! \, \mathchar`- \, ?c$)}

    \caption{\takyurevise{Principal} cut elimination rules of MELL.}\label{figure:LL-cut-full}
\end{figure}

\subsection{Proof Structure and Proof Nets}\label{subsection: proofnets}
We define \textit{proof nets} 
corresponding to the
sequent proofs of MELL 
with mix rules,
following the style of \cite{lionel_tutorial}.
First, we define \textit{proof structures}. 

\begin{definition}%
\label{def: ps}
  An \emph{MELL proof structure}
  is a directed acyclic multigraph
  that combines the cells (not to be confused with cells 
  $m\textup{\lmem}P\textup{\rmem}$
  of
  LMNtal)
  and wires shown 
  in \cref{figure: ps}(a) and the
  promotion box
  shown 
  in \cref{figure: ps}(b). 
\end{definition}

Each component of \cref{figure: ps}(a) 
consists of
\begin{enumerate}
  \item cells (nodes) labeled with MELL inference rules and
  \item wires (edges) labeled with MELL formulas.
\end{enumerate}    
The cells $\otimes$ and $\linpar$ have two \textit{ordered} inputs,
while the inputs of other cells are \textit{unordered.}

The inference rule for the bottom-right component of \cref{figure: ps}(a),
called \textit{promotion}, is a 
\textit{contextual rule};
that is, the rest of the formulas of the sequent containing the $!A$
must come with `$?$'s.
To handle this constraint, 
a structure called a 
\textit{promotion box} (\cref{figure: ps}(b), also simply called a 
\textit{box})
is used as a standard mechanism
to protect its contents
from rewriting and 
control the order of
proof reductions \cite{girard_linear_1987}.
Here, the conclusion with `$!$' is called the \textit{principal door} of
the box, while the (bundle of) conclusions with `$?$' are called the
\textit{auxiliary doors}.
%

A proof structure does not necessarily
correspond to a
proof tree, but proof structures satisfying a ``correctness criterion''
have 
corresponding
proofs of sequent calculus
and are
called \textit{proof nets}. 
There are several known correctness criteria that are equivalent,
but 
we adopt the popular Danos-Regnier correctness   
\cite{danos_structure_1989}
based on
  \textit{switching graphs} 
formed by
non-deterministically 
\ueda{choosing}
and cutting
either the left or 
right input of the $\linpar$ and $?c$ cells. 
%

\begin{definition}%
\label{def: sw}
 A \emph{switching} 
 of a proof structure
 is the choice of 
the left or the right input for each $\linpar$ or $?c$,
as shown in \cref{figure: sw}.
 A \emph{switching graph} 
 is obtained by replacing each $\linpar$ or $?c$
 \takyurevise{(\cref{figure: sw}(a))}
 accordingly,
 \takyurevise{and by replacing each box with a single \emph{initial
 node} $i$ (\cref{figure: sw}(b)).
 Here, an \emph{initial node} is a node that has no premise and all of
 whose links are conclusions
\ueda{(such as the $ax$ cell).}
 The conclusions of the node $i$ that replaces a box are the doors of
 that box, namely its principal door `$!A$' and its auxiliary doors
 `$?\Gamma$'}.
\end{definition}
\begin{definition}%
\label{def: pn}
  A 
  \emph{proof net}
   is a proof structure whose switching graphs 
  have no undirected cycles
  and such that the content of each box is a proof net,
    inductively.

\end{definition}

\takyurevise{%
Note that, since a switching graph replaces each box with a single
initial node, the content of a box does not contribute to the undirected
cycles of the switching graph; it is instead checked separately by the
second, inductive condition above.}

\begin{figure}[t]
  \begin{minipage}[t]{0.49\hsize}
      \begin{center}
        \scalebox{0.65}{\input{figures/3/ps.tex}}\\
      (a) Cells and Wires
      \end{center}
  \end{minipage}
  \hfill
  \begin{minipage}[t]{0.5\hsize}
      \begin{center}
        \raisebox{-16pt}{\scalebox{0.65}{\input{figures/3/box_inf.tex}}}\\[-14pt]
      (b) Promotion box for the promotion rule
     \end{center}
  \end{minipage}
    \caption{The components of MELL proof structure.}
    \vspace{-0.75em}
    \label{figure: ps}
\end{figure}

\begin{figure}[t]
    \begin{center}
        \fontsize{14pt}{14pt}\selectfont
        \begin{tikzpicture}
          \node (par) at (0,0) {\scalebox{0.45}{\input{figures/3/switching.tex}}};
          \node (cont) at (4,0) {\scalebox{0.45}{\input{figures/3/switching_cont.tex}}};
          \node (a) at (1.9,-1.5) {\normalsize(a)};
          \node (box) at (8,0) {\scalebox{0.45}{\input{figures/3/switching_box.tex}}};
          \node (b) at (8.1,-1.5) {\normalsize(b)};
        \end{tikzpicture}
    \end{center}
    \vspace{-1.5em}
    \caption{\ueda{Construction of switching graphs.}}
  \label{figure: sw}
\end{figure}


\begin{figure}[t]
    \newlength{\lamheight}
    \settoheight{\lamheight}{\scalebox{0.4}{\input{figures/3/before.tex}}}
    \begin{minipage}[t]{0.82\hsize}
    \begin{center}
        \scalebox{0.45}{\input{figures/3/before.tex}}\\[-8pt]
        (a)
    \end{center}
    \end{minipage}
    \hfill
    \begin{minipage}[t]{0.17\hsize}
    \begin{center}
        \raisebox{0.3\lamheight}%
        {\scalebox{0.45}{\input{figures/3/after.tex}}}\\[-12pt]
        (b)
    \end{center}
    \end{minipage}
  \caption{Examples of MELL proof nets,
  where applying cut elimination to (a) results in (b)
 (See \cref{subsection: example-beta-reduction} also).}
  \label{figure: pn}
\end{figure}

\cref{figure: pn} shows examples of MELL proof nets.
Since 
every undirected cycle
contains a $\linpar$, and
all the switching graphs
have no undirected cycles,
the Danos-Regnier correctness
is satisfied. 
The proof tree corresponding to \cref{figure: pn}(a)
will be given in \cref{figure: pt}.

\subsection{Cut Elimination}

Cut elimination of MELL proof nets is expressed as graph rewriting rules.
%
%

\cref{figure: pn-cut-full} shows all the cut elimination rules in MELL proof nets,
involving (c) migration, (e) deletion, and (f) cloning of boxes, whose 
concise representation in a graph rewriting language is the topic
of the next sections.


\begin{figure}[t]
    \begin{minipage}[t]{0.45\textwidth}
    \begin{center}
        \scalebox{0.5}{\input{figures/3/cut1}}

        (a) Rule (\textit{ax}-\textit{cut})
    \end{center}
    \end{minipage}
    \begin{minipage}[t]{0.45\textwidth}
    \begin{center}
        \scalebox{0.5}{\input{figures/3/cut2}}

        (b) Rule ($\otimes \, \mathchar`- \, \linpar$)
    \end{center}
  \end{minipage}


    \begin{center}
        \scalebox{0.5}{\input{figures/3/cut-nest.tex}}
        \scalebox{0.5}{\input{figures/3/cut-nest2.tex}}

        (c) Rule ($! \, \mathchar`- \, !$)
    \end{center}

  \begin{minipage}[t]{0.45\textwidth}
    \begin{center}
        \scalebox{0.5}{\input{figures/3/cut-der}}

        (d) Rule ($! \, \mathchar`- \, ?d$)
    \end{center}
  \end{minipage}
  \begin{minipage}[t]{0.45\textwidth}
    \begin{center}
        \scalebox{0.5}{\input{figures/3/cut-weakn.tex}}

        (e) Rule ($! \, \mathchar`- \, ?w$)
    \end{center}
  \end{minipage}

    
    \begin{center}
        \scalebox{0.5}{\input{figures/3/cut-cont1.tex}}
        \scalebox{0.5}{\input{figures/3/cut-cont2.tex}}

        (f) Rule ($! \, \mathchar`- \, ?c$)
    \end{center}

  \caption{Cut elimination rules in MELL proof nets.}
  \label{figure: pn-cut-full}
\end{figure}

%

The following
hold for cut elimination of MELL proof nets
\cite{girard_linear_1987,girard_linear_1993,pagani_strong_2010}:

\begin{enumerate}
\item \textbf{(Cut Elimination)}
  All cuts of an MELL proof net can be eliminated.

\item \textbf{(Stability)}
  An MELL proof net is still a
  proof net after cut elimination.

\item \textbf{(Confluence)}
    Cut elimination is confluent on MELL proof nets.

\item \textbf{(Strong Normalization)}
    Cut elimination is strongly normalizing.
\end{enumerate}


\section{Refining LMNtal's Box Constructs}\label{section: ext}

\begin{figure}[t]
  \begin{center}
  \begin{tikzpicture}
    \node (be) at (0,0) {\scalebox{0.57}{\input{figures/4/cp.tex}}};
    \node (a1) at (4,0) {\scalebox{0.57}{\input{figures/4/cp1.tex}}};

    \draw[->,thick] (0.75,0) -- (2.8,0);
    \node at (1.5,-0.5) {Copy};
    \draw[thick,circle,red] (2.3,-0.5) circle [radius=0.2];
    \node at (2.3,-0.5) {$b$};
  \end{tikzpicture}
  \end{center}
  \vspace*{-1em}
  \caption{Cloning of a subgraph.}
  \label{figure: cp_prob}
\end{figure}

A design challenge of a practical declarative language for graph
rewriting is the design of high-level constructs for the manipulation
of subgraphs of non-fixed shape and size
(as opposed to individual elements).

For example, consider the cloning of a subgraph containing the node \texttt{b} as shown in \cref{figure: cp_prob}.
After cloning, not only the node \texttt{b} but also the two edges of the node \texttt{b} are cloned.
However, whether or not the two new edges can be connected to \texttt{a}
and \texttt{c} depends on the definition of the graph. 
In LMNtal which handles port graphs consisting of atoms with
fixed arities, 
\texttt{a} and \texttt{c} has no ports for the new edges.
We must somehow splice two clones of the subgraph into the context of the original subgraph.
%
This can be divided into \takyurevise{:} 
\begin{enumerate}
\item how to specify subgraphs for rewriting, and
\item how to 
copy, delete, or migrate them in a single rewriting step.
\end{enumerate}
One direction was studied by \cite{Alim-Access} to find the right notion of
\textit{ground graphs} (a graph counterpart of \textit{ground terms})
for HyperLMNtal, 
with applications to the encoding of formal systems with name binding,
but it
focused on \textit{operations based on connectivity}. 
We 
focus on the manipulation of subgraphs delimited by membranes,
that is, \textit{operations based on hierarchy}.

An approximate solution using bundles and aggregates was proposed
(\cref{section: lmntal}), and the \texttt{nlmem} API
(\cref{subsection: nlmem}) has already been put in practice.
This included our initial attempt of encoding MELL proof nets
(\cref{section: intro}) using \texttt{nlmem} 
which was designed totally
independently of proof nets.
However, we found that the solution was only approximate.  Whereas the
functionality of \texttt{nlmem} (\cref{figure: nlmem}(a)(b)) looks
similar to that of promotion box operations
(\cref{figure: pn-cut-full}(f)(e), respectively), 
a refinement is necessary to
allow straightforward encoding because:
\begin{enumerate}
\item 
  the functionality did not distinguish between
  principal and auxiliary doors
  of boxes, requiring
  post-processing to obtain the desired result, and
\item while the processing of (the free links of) bundles was the key design
  issue, the only functionality provided was to reconfigure them with
  the collection of atoms, which turned out to be restrictive.
\end{enumerate}

\subsection{Extension of LMNtal Syntax: Aggregates of Process Contexts}

We propose an extension of LMNtal that solves the
above issues and provides constructs for cloning and deleting
membranes with indefinitely many free links. 

%
\takyurevise{
Recall from \cref{section: lmntal_overview} that an aggregate of atoms
$p(\texttt{*}X_1,\ldots,\texttt{*}X_n)$ 
\ueda{stands for}
a multiset of
atoms named $p$ whose multiplicity matches the size of each argument
bundle, and that a \emph{process context} $\texttt{\$}p[\ldots]$ is a
wildcard matching a subgraph within a membrane.  We extend the
syntax of process templates to lift 
\ueda{the notion of}
aggregates from atoms to process
contexts:}
\[
  \begin{array}{cccccc}
T &::=&  \dots &~\bigm|~& \texttt{\$}p\texttt{[}\texttt{*}X_1\pc \dots \pc \texttt{*}X_n\texttt{]}&\quad (n>0) \\
  \end{array}
\]
where each $\texttt{*}X_i$ is a bundle that 
\ueda{occurs}
in the process
context with the same name, 
which implies that
$|\texttt{*}X_1| = \dots = |\texttt{*}X_n |$,
where $|\texttt{*}X_i |$ stands for the number of links 
represented
by
$\texttt{*}X_i$ upon the present rewriting.
%
%
\takyurevise{
Whereas an aggregate of atoms 
\ueda{stands for}
a multiset of atoms (each of a fixed
arity $n$), an aggregate of process contexts 
\ueda{stands for}
$|\texttt{*}X_i|$
copies of $\texttt{\$}p$, each 
with $n$ free links
drawn 
\ueda{from each of $\texttt{*}X_1,\dots,\texttt{*}X_n$.}}
\ueda{(A concrete example will appear immediately below 
in \cref{section:copy}.)}
%
%
As a practical decision, this functionality is provided
as an API as was the case of $\texttt{nlmem}$.

\subsection{API Rules Supporting Typical Use Cases}

\subsubsection{Copying a membrane with a bundle of free links}\label{section:copy}
We first give a construct for copying a membrane with an
indefinite number of free links, which can be specified as follows
and illustrated in \cref{figure: mell}(a):

\begin{specbox}\small
    \texttt{mell.copy({\color{red}{X}},
      {\color{orange}{A1}},{\color{violet}{A2}},{\color{blue!80}{A3}}, {\color{red}{B1}},{\color{green2}{B2}}, {\color{green2}{C1}},{\color{green2}{C2}}),\\
  \texttt{\{\$p[{\color{red}{X}}|{\color{blue!80}{*Z}}]\}}, \{\$a[{\color{orange}{A1}},{\color{violet}{A2}},{\color{blue!80}{A3}}]\}, \{\$b[{\color{red}{B1}},{\color{green2}{B2}}]\}}\\
  \hspace{10em} $\react$\;\;\texttt{\{\$p[{\color{red}{X'}}|{\color{orange}{*Z'}}]\}, \{\$p[{\color{red}{X''}}|{\color{violet}{*Z''}}]\}},\\
  \hspace{12em} $\texttt{\$a[{\color{orange}{*Z'}}\pc {\color{violet}{*Z''}}\pc{\color{blue!80}{*Z}}], }$\texttt{\$b[{\color{red}{X'}},{\color{green2}{C1}}], \$b[{\color{red}{X''}},{\color{green2}{C2}}].}
\end{specbox}

\noindent 
The first argument $\texttt{X}$ of $\texttt{mell.copy}$ 
(where $\texttt{mell}$ is the prefixed module name)
is linked to 
the membrane to be copied
\ueda{(\cref{figure: mell}(a)(left)}).
We call the occurrence of $\texttt{X}$ in $\texttt{\$p[X|*Z]}$
the \textit{principal port} of $\texttt{\$p}$.
The next five arguments
(\texttt{A1}, \texttt{A2}, \texttt{A3}, \texttt{B1}, \texttt{B2})
are 
linked
to structures (\texttt{\$a[A1,A2,A3]} and \texttt{\$b[B1,B2]})
to be connected to the
free links of the copies of \texttt{\$p}.
The three process contexts are enclosed by membranes because
the subgraphs matched by the process contexts are 
determined by their
delimiting membranes.
The final two arguments \texttt{C1} and \texttt{C2} are links that
will be connected to the two copies of \texttt{\$p}.

The arguments
\texttt{A1}, \texttt{A2}, \texttt{A3} 
are linked to
\texttt{\$a} for handling copies of the links \takyurevise{$\texttt{*Z}$} connected
to \texttt{\$p}'s non-principal ports.
The 
arguments \texttt{B1} and \texttt{B2}
are linked
to \texttt{\$b} for handling copies of
\texttt{X} connected to \texttt{\$p}'s principal port.
The 
arguments \texttt{C1} and \texttt{C2}
will be linked to
the principal ports
of the two copies of \texttt{\$p} \textit{via copies of} \texttt{\$b}.
After the reduction, the structure $\texttt{\{\$p[X|*Z]\}}$ 
is cloned,
generating $\texttt{\{\$p[X'|*Z']\}}$ and $\texttt{\{\$p[X''|*Z'']\}}$
\ueda{(\cref{figure: mell}(a)(right))}.
Each of the process contexts obtained by cloning $\texttt{\$a}$
connects a member of $\texttt{*Z'}$ and 
a member of $\texttt{*Z''}$ 
to
a member of the original bundle $\texttt{*Z}$. 
The two copies of
$\texttt{\$b}$ connect $\texttt{X'}$ and $\texttt{X''}$ of the
principal ports 
of $\texttt{\$p}$ to $\texttt{C1}$ and $\texttt{C2}$,
  respectively.
Thus, \texttt{mell.copy} separates the treatment of the principal port
from that of non-principal ports.

\subsubsection{Deleting a membrane with a bundle of free links}

Next, we present a construct for deleting a membrane with an
unspecified number of free links, which is defined as follows and
illustrated in \cref{figure: mell}(b):

\begin{figure}[t]
     \begin{center}
           \newlength{\copyheightt}
           \settoheight{\copyheightt}{\scalebox{0.7}{\input{figures/4/copy-after.tex}}}
           \raisebox{0.05\copyheightt}{\scalebox{0.7}{\input{figures/4/copy-before.tex}}}
           \raisebox{0.45\copyheightt}{\LARGE${}~~~\to~~~{}$}
           \scalebox{0.7}{\input{figures/4/copy-after.tex}}\\
    (a) \texttt{mell.copy}
     \end{center}
     \medskip
      \begin{center}
           \raisebox{4mm}{\scalebox{0.7}{\input{figures/4/delete-before.tex}}}
           \raisebox{12mm}{\LARGE${}~~~\to~~~{}$}
           \raisebox{3mm}{\scalebox{0.7}{\input{figures/4/delete-after.tex}}}\\
    (b) \texttt{mell.delete}
      \end{center}

  \caption{Schematic illustrations of the \texttt{mell} library.
} 
  \label{figure: mell}
\end{figure}

\begin{specbox}\small
      \texttt{mell.delete({\color{red}{X}},{\color{blue!80}{A}}), \{\$p[{\color{red}{X}}|{\color{blue!80}{*X}}]\}, \{\$a[{\color{blue!80}{A}}]\}} $\react$ $\texttt{\$a[{\color{blue!80}{*X}}].}$
\end{specbox}

\noindent
This is  a generalization of
\texttt{nlmem.kill} to allow non-atomic structures
to terminate the members of \texttt{*X} (which would otherwise become
dangling links).

\subsubsection{Implementation}

The \texttt{mell} library was implemented in approximately 400 LOC of C++ within the LMNtal runtime. 
The source code has already been merged into the develop branch of SLIM. 
We have confirmed that all the examples%
\footnote{%
  The complete source code used in the examples 
  in \cref{section: encoding-pn} and \cref{section: encoding-proc} can be found in 
  \url{https://lmntal.github.io/mell-library-examples}.
} 
in the following sections work.

\section{Encoding of MELL Proof Nets and Cut Elimination Rules}\label{section: encoding-pn}
This section shows how 
we can (i) encode 
the cut elimination rules 
of MELL proof nets 
using the constructs introduced in \cref{section: ext} and
(ii) use the model checker of LMNtal to verify
various properties of proof nets 
to demonstrate
that LMNtal serves as a useful workbench for proof nets.

\begin{figure}[t]
    \begin{center}
        \scalebox{0.65}{\input{figures/5/ps1.tex}}
        \hspace{-.5em}\scalebox{0.65}{\input{figures/5/ps2.tex}}\\
        \scalebox{0.65}{\input{figures/5/ps-lmn1.tex}}
        \hspace{-.7em}\scalebox{0.65}{\input{figures/5/ps-lmn2.tex}}
    \end{center}
    \vspace{-.5em}
  \caption{Encoding of cells and wires
    (MELL in {\color{red} red}, LMNtal in {\color{blue} blue} henceforth);
    where a small circle in a membrane stands for a unary atom named `\texttt{+}'
    and may be written as a prefix operator in textual representation.}
\label{figure: encodeps}
\end{figure}

\subsection{MELL Proof Nets}

First, we encode the components of the Proof Structure (\cref{figure: ps}).
Figure~\ref{figure: encodeps} shows the encoding of cells and wires.
The 
non-commutativity
of inputs to $\otimes$ and $\linpar$ cells is
represented by atoms, while
the 
commutativity
of the arguments of \texttt{ax} and \texttt{cut}
is represented using membranes. 
A $?c$ cell with 
commutative
inputs and a single output is
represented by using both an atom and a membrane.

\begin{figure}[t]
		\centering
        \scalebox{0.6}{\input{figures/5/box.tex}}
        \scalebox{0.6}{\input{figures/5/box_mem.tex}}
  \caption{Encoding of a promotion box.}\label{figure: encodepb}
\end{figure}

Figure~\ref{figure: encodepb} shows the encoding of a promotion box
(\cref{figure: ps}(b)).
The outer frame of the promotion box is represented by a membrane.
The context $?\Gamma$ is represented by a bundle $\texttt{*X}$, and the blank part is represented by a process context $\texttt{\$p[X1|*X]}$.


As described above, all the components of a proof structure can be
encoded directly, and 
the entire proof structure can be encoded by
combining these encodings. 

Figure~\ref{figure: pt} shows the proof tree corresponding to \cref{figure: pn} and its LMNtal encoding.

\begin{figure}[t]
     \tiny
        \begin{center}
          \[
            \hspace{-2em}
            \begin{prooftree}
              \hypo{}
              \infer1[($ax$)]{\vdash !n \otimes n^{\bot}, ?n^{\bot} \linpar n}
              \infer1[($?d$)]{\vdash ?(!n \otimes n^{\bot}), ?n^{\bot} \linpar n}
              \infer1[($?w$)]{\vdash ?(!n \otimes n^{\bot}), ?n^{\bot}, ?n^{\bot} \linpar n}
            
              \hypo{}
              \infer1[($ax$)]{\vdash n^{\bot}, n}
              \infer1[($?d$)]{\vdash ?n^{\bot}, n}
              \infer1[($?w$)]{\vdash ?(!n \otimes n^{\bot}), ?n^{\bot}, n}
              \infer1[($!$)]{\vdash ?(!n \otimes n^{\bot}), ?n^{\bot}, !n}
            
              \hypo{}
              \infer1[($ax$)]{\vdash n, n^{\bot}}
              \infer2[($\otimes$)]{\vdash ?(!n \otimes n^{\bot}), ?n^{\bot}, n, !n \otimes n^{\bot}}
            
              \infer2[($cut$)]{\vdash ?(!n \otimes n^{\bot}), ?(!n \otimes n^{\bot}), ?n^{\bot}, ?n^{\bot}, n}
              \infer1[($?c$)]{\vdash ?(!n \otimes n^{\bot}), ?(!n \otimes n^{\bot}), ?n^{\bot}, n}
              \infer1[($?c$)]{\vdash ?(!n \otimes n^{\bot}), ?n^{\bot}, n}
              \infer1[($\linpar$)]{\vdash ?(!n \otimes n^{\bot}), ?n^{\bot} \linpar n}
              \infer1[($\linpar$)]{\vdash ?(!n \otimes n^{\bot}) \linpar ?n^{\bot} \linpar n}
            
              \hypo{}
              \infer1[($ax$)]{\vdash n^{\bot}, n}
              \infer1[($?d$)]{\vdash ?n^{\bot}, n}
              \infer1[($\linpar$)]{\vdash ?n^{\bot} \linpar n}
              \infer1[($!$)]{\vdash !(?n^{\bot} \linpar n)}
            
              \hypo{}
              \infer1[($ax$)]{\vdash !n \otimes n^{\bot}, ?n^{\bot} \linpar n}
              \infer2[($\otimes$)]{\vdash !(?n^{\bot} \linpar n) \otimes (!n \otimes n^{\bot}), ?n^{\bot} \linpar n}
            
              \infer2[($cut$)]{\vdash ?n^{\bot} \linpar n}
            \end{prooftree}
          \]
        \end{center}
        \normalsize
        \[
          \Large
          \downmapsto
        \]
    \begin{center}
        \scalebox{0.45}{\input{figures/3/before.tex}}
    \end{center}
  \vspace*{-18pt}
        \[
          \Large
          \downmapsto
        \]
  \vspace*{-14pt}
  \begin{lstlisting}[basicstyle=\ttfamily\footnotesize]
  ax{+A1,+A2}, '?d'(A1,A3), '?w'(A4),
  {ax{+B1,+B2}, '?d'(B1,B3), '?w'(B4), '!'(B2,B5)},
  '?c'({+A3,+B4},C1), '?c'({+A4,+B3},C2),
  tensor(B5,T1,D2), ax{+T1,+T2}, cut{+A2,+D2}, par(C2,T2,P1), par(C1,P1,F),
  {ax{+E1,+E2}, '?d'(E1,E3), par(E3,E2,E4), '!'(E4,E5)},
  tensor(E5,T3,D4), ax{+T3,+T4}, cut{+F,+D4},
  formula(T4).
\end{lstlisting}
\vspace*{-0.5em}
    \caption{Proof tree corresponding to \cref{figure: pn}(a) and its
      LMNtal encoding.}\label{figure: pt} 
\end{figure}



\subsection{Cut Elimination Rules}

We encode the cut elimination rules (\cref{figure: pn-cut-full})
into the rewrite rules of LMNtal.
%

Rule ($! \, \mathchar`- \, !$) is encoded as follows
(as noted in \cref{section: lmntal_overview},
each LMNtal rule may be given an optional rule name
followed by \texttt{@@}).

\begin{lstlisting}[basicstyle=\ttfamily\small]
  promotion_promotion@@
  {'!'(X1,X2), $p[X1|*X]}, cut{+X2,+X3}, {$q[X3,X4|*Y], '!'(X4,X5)}
    :- {{'!'(X1,X2), $p[X1|*X]}, cut{+X2,+X3}, $q[X3,X4|*Y], '!'(X4,X5)}.
\end{lstlisting}
Figure~\ref{figure: encodenest} shows the corresponding illustration.
We can see that the encoding using LMNtal's bundles and membranes
is quite straightforward.
%
Furthermore, under the Occurrence Conditions of bundles
\cite{ueda_lmntal_2009}, 
the rewrite rule is guaranteed 
not to generate dangling links.

\begin{figure}[t]
    \newlength{\nheight}
    \settoheight{\nheight}{\scalebox{0.48}{\input{figures/5/cut-nest.tex}}}
    \begin{center}
        \scalebox{0.37}{\input{figures/5/cut-nest}}
        \scalebox{0.37}{\input{figures/5/cut-nest2}}
        \raisebox{0.5\nheight}{\Large$\mapsto$}
        \scalebox{0.37}{\input{figures/5/cut-nest-lmn.tex}}
        \scalebox{0.37}{\input{figures/5/cut-nest2-lmn.tex}}
    \end{center}
    \vspace{-1em}
  \caption{Encoding of Rule ($! \, \mathchar`- \, !$).}
  \label{figure: encodenest}
\end{figure}

\begin{figure}[t]
    \newlength{\bwheight}
    \settoheight{\bwheight}{\scalebox{0.48}{\input{figures/5/cut-weakn.tex}}}
    \begin{center}
      \scalebox{0.45}{\input{figures/5/cut-weakn.tex}}
      \hspace{0.4em}
      \raisebox{0.5\bwheight}{\LARGE$\mapsto$}
      \hspace{0.4em}
      \scalebox{0.45}{\input{figures/5/cut-weakn-lmn.tex}}
      \scalebox{0.45}{\input{figures/5/cut-weakn-mid-lmn.tex}}
      \scalebox{0.45}{\input{figures/5/cut-weakn-after-lmn.tex}}
    \end{center}
    \begin{flushleft}
        \scalebox{0.45}{\input{figures/5/cut-cont1}}
        \scalebox{0.45}{\input{figures/5/cut-cont2}}
       \raisebox{0.75\bwheight}{\LARGE$\mapsto$}
    \end{flushleft}
    \vspace{-1.2em}
    \begin{flushright}
      \scalebox{0.45}{\input{figures/5/cut-cont1-lmn.tex}}
      \scalebox{0.45}{\input{figures/5/cut-cont-new-mid12-lmn.tex}}
      \scalebox{0.45}{\input{figures/5/cut-cont2-lmn.tex}}
    \end{flushright}
  \caption{Encoding of Rule ($! \, \mathchar`- \, ?w$) and 
Rule ($! \, \mathchar`- \, ?c$) with the $\texttt{mell}$ library.}
  \label{figure: encodemell}
\end{figure}

\begin{figure}[t]
  \begin{lstlisting}[basicstyle=\ttfamily\small]
  ax_cut@@
  cut{+X,+Y}, ax{+Y,+Z} 
    :- X=Z.

  tensor_par@@
  tensor(X1,Y1,XY1), par(X2,Y2,XY2), cut{+XY1,+XY2} 
    :- cut{+X1,+X2}, cut{+Y1,+Y2}.

  promotion_promotion@@       // !-!
  {'!'(X1,X2), $p[X1|*X]}, cut{+X2,+X3}, {$q[X3,X4|*Y], '!'(X4,X5)}
    :- {{'!'(X1,X2), $p[X1|*X]}, cut{+X2,+X3}, $q[X3,X4|*Y], '!'(X4,X5)}.

  promotion_dereliction@@    // !-?d
  {'!'(X1,X2), $p[X1|*X]}, cut{+X2,+X3}, '?d'(X4,X3)
    :- cut{+X1,+X4}, $p[X1|*X].

  promotion_weakening@@      // !-?w
  {'!'(X1,X2), $p[X1|*X]}, cut{+X2,+X3}, '?w'(X3) 
    :- mell.delete(X1,W), {$p[X1|*X]}, {'?w'(W)}.

  promotion_contraction@@    // !-?c
  {'!'(X1,X2), $p[X1|*X]}, cut{+X2,+X3}, '?c'({+C1,+C2}, X3)
    :- mell.copy(X2,A1,A2,A3,B1,B2,C1,C2),{'!'(X1,X2),$p[X1|*X]},
       {'?c'({+A1,+A2}, A3)}, {cut{+B1,+B2}}.
  \end{lstlisting}
  \vspace*{-0.5em}
  \caption{LMNtal encoding of cut elimination rules (full).}
  \label{code: full}
\end{figure}

Other cut elimination rules can be encoded in a similarly concise
manner 
(all the encodings are given in \cref{code: full}), 
of which the
weakening and contraction rules involve non-trivial operations and
need to be encoded using the constructs introduced in
\cref{section: ext}, as described below.

Rule ($! \, \mathchar`- \, ?w$) is encoded as follows.

\needspace{7\baselineskip}
\begin{lstlisting}[basicstyle=\ttfamily\small]
  promotion_weakening@@
  {'!'(X1,X2), $p[X1|*X]}, cut{+X2,+X3}, '?w'(X3) 
    :- mell.delete(X,A), {$p[X|*X]}, {'?w'(A)}.
\end{lstlisting}
Figure~\ref{figure: encodemell} (upper) shows the corresponding
illustration. 
By using $\texttt{mell.delete}$, the $?w$ atoms are connected to the links
previously connected to the non-principal ports of the 
promotion box.
Note that the encoded rewriting is done in two steps due to the API
call.


Rule ($! \, \mathchar`- \, ?c$) is encoded as follows.
\begin{lstlisting}[basicstyle=\ttfamily\small]
  promotion_contraction@@
  {'!'(X1,X2), $p[X1|*X]}, '?c'(I,X3), {+I,+C1,+C2}, cut{+X2,+X3},
    :- mell.copy(;\color{red}{X2};, ;\color{orange}{A1};,;\color{violet}{A2};,;\color{blue!80}{A3};, ;\color{red}{B1};,;\color{green2}{B2};, ;\color{green2}{C1};,;\color{green2}{C2};),
       {'!'(X1,X2), $p[X1|*X]},
       {'?c'(I, ;\color{blue!80}{A3};), {+I,+;\color{orange}{A1};,+;\color{violet}{A2};}}, {cut{+;\color{red}{B1};,+;\color{green2}{B2};}}.
\end{lstlisting}
Figure~\ref{figure: encodemell} (lower) shows the corresponding
illustration.  
The second to fourth arguments of $\texttt{mell.copy}$ 
in \cref{figure: encodemell}(lower middle)
are linked to
the structure
corresponding to the $?c$ cell. 
The main reason for allowing non-atomic structures 
in $\texttt{mell.copy}$ 
lies in this encoding.
\ueda{Its} fifth and sixth arguments correspond to the $cut$ cell.
The seventh and eighth arguments are linked to
the input side of the
contraction before cut elimination to build
the structure around the principal port after cut
elimination. 
With $\texttt{nlmem.copy}$, it was necessary to use multiple rules 
and unbounded reduction steps to
reconfigure
free links 
using non-atomic structures 
and to handle the
principal
port, but with the introduction of $\texttt{mell.copy}$, it is
now possible to encode it with a single rule. 

\subsection{Correctness of the Encoding}

Regarding the correctness of the encoding,
we note that Figs.~\ref{figure: encodeps}--\ref{figure: encodemell}
show clear correspondence between proof nets with cut
elimination rules and their LMNtal encodings,
which indeed is the main purpose of the present work.
When a formal proof is required, it is necessary to establish a
correspondence between the mathematical representation of each
side, but we omit the details in this paper. 




\subsection{Example: State space of $\beta$-reduction by cut elimination}
\label{subsection: example-beta-reduction}

\begin{figure}[t]
      \small
      \[
      \begin{prooftree}[rule margin=0.4ex]
        \hypo{}
        \infer1[$\text{T-Var}$]{f: n \to n,\, x : n \vdash f : n \to n}
        \hypo{}
        \infer1[$\text{T-Var}$]{f: n \to n,\, x : n \vdash x  : n}
        \infer2[$\text{T-App}$]{f: n \to n,\, x : n \vdash f x : n}
        \infer1[$\text{T-Abs}$]{f: n \to n \vdash \lambda x : n\,.\, f x : n \to n}
        \infer1[$\text{T-Abs}$]{\vdash \lambda f: n \to n\,.\, \lambda x : n\,.\, f x : n \to n \to n \to n}
        
        \hypo{}
        \infer1[$\text{T-Var}$]{x: n\vdash x:n\vphantom{f}}
        \infer1[$\text{T-Abs}$]{\vdash \lambda x : n\,.\, x : n \to n}
        \infer2[$\text{T-App}$]{\vdash (\lambda f : n\to n\,.\,\lambda x : n\,.\,fx)(\lambda x : n . x) : n \to n}
      \end{prooftree}
      \] 
      \[
        \quad \rightsquigarrow^* \quad
        \begin{prooftree}[rule margin=0.4ex]
          \infer0[$\text{T-Var}$]{x : n \vdash x : n\vphantom{f}}
          \infer1[$\text{T-Abs}$]{\vdash \lambda x : n\,.\, x : n \to n}
        \end{prooftree}
      \]
  \caption{\ueda{Proof trees corresponding to}
      $(\lambda f \,\mathord{:}\, n \to n \:.\: 
        \lambda x \,\mathord{:}\, n \: . \: f \: x)\:
        (\lambda x \, \mathord{:} \, n \: .\: x )$
       and
      $(\lambda x \, \mathord{:}\, n \: . \: x)$.}\label{table:lam}
\end{figure}

\begin{figure}[t]
	\centering
       \begin{center}
    \begin{tikzpicture}

    \node (fig) at (6.4,0) {\includegraphics[width=0.75\textwidth]{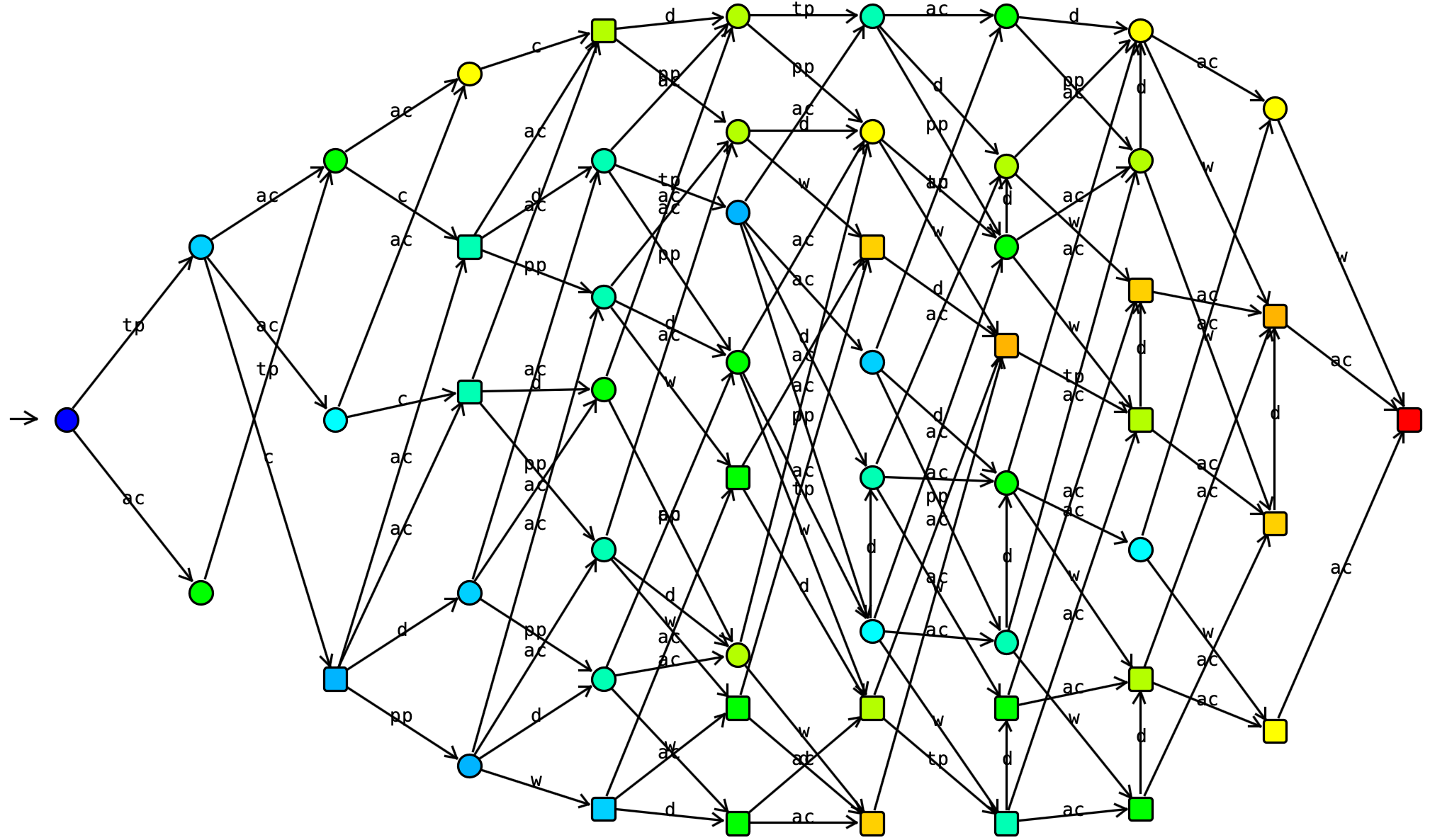}};

    \node[draw] (Before) at (1.3,-2.8) {\scalebox{0.2}{\input{figures/3/before.tex}}};
    \node[draw] (After) at (11.0,-3) {\scalebox{0.2}{\input{figures/3/after.tex}}};

    \node[circle,draw,densely dotted,minimum size=0.5em] (initstate) at (2.05,0) {};
    \node[circle,draw,densely dotted,minimum size=0.5em] (endstate) at (11.06,0) {};

    \draw[densely dotted] (initstate.210) -- (Before.north west);
    \draw[densely dotted] (initstate.330) -- (Before.north east);
    \draw[densely dotted] (endstate.255) -- (After.north west);
    \draw[densely dotted] (endstate.285) -- (After.north east);

    \end{tikzpicture}
  \end{center}

    \caption{State space of cut elimination applied to the proof net of
\cref{figure: pn}. \takyurevise{
\ueda{The leftmost (blue) node and the rightmost (red) node
mark the initial and final states, which are
magnified into the boxed nets on the bottom left and right, respectively.
The initial state
is the 
proof net of \cref{figure: pn}(a), while the final state is
the proof net of \cref{figure: pn}(b) reached after
eliminating all cuts.}}}\label{figure: stateex}
\end{figure}

We represent simply typed $\lambda$-calculus in proof nets.
By $(A \to B) \:\mapsto \: (?A^{\bot} \:\linpar\: B)$, 
the typing of 
$\lambda$-terms
can be embedded into MELL proof nets
\cite{girard_linear_1987,guerrini_proof_2004}. 
As is well-known,
cut elimination 
refines
$\beta$-reduction, and the proof net
after eliminating all cuts corresponds to the normal form of the
$\lambda$-term.
Figure~\ref{figure: pn}(a) \ueda{(\cref{subsection: proofnets})}
shows the proof-net representation of the 
typed $\lambda$-term 
\ueda{%
$(\lambda f\,\mathord{:}\, n\mathord{\to} n \: . \:
\lambda x \,\mathord{:}\, n \: . \: f\: x)\:(\lambda x \,\mathord{:}\,
n \: . \: x)$,}
which corresponds to the proof tree of \cref{table:lam} (upper). 
\ueda{This $\lambda$-term}
is reduced by $\beta$-reduction to the normal form $\lambda x\,
\mathord{:}\, n \: .\: x$,
%
whose corresponding proof tree
is in \cref{table:lam} (lower), and the 
corresponding net is the proof net of \cref{figure: pn}(b). 

Figure~\ref{figure: stateex} shows the state space of
the cut elimination 
of this example
visualized by LaViT (LMNtal Visual Tool).
Note that each rule using the \texttt{mell} library involves one extra
reduction step
(\cref{figure: encodemell}),
but Transition Abstraction of
LaViT allows one to 
visualize the states before and after
their application as a single abstract state
(shown as squares 
rather than circles
in \cref{figure: stateex}).
We can see from \cref{figure: stateex} that
the initial net is reduced to the net corresponding to the
normal form $\lambda x \,\mathord{:}\, n \: . \: x$ in 
10 or more steps. 
Furthermore, 
since all reduction paths eventually stop at a single final state
(the red node in the figure),
we can easily observe the confluence
and strong normalization 
for this example.

\subsection{Example: Adding Push-equivalence Rules}\label{subsec: push-eq}

One advantage of the direct encoding of 
proof nets 
into LMNtal
is that 
one can easily
modify and/or add rewrite rules and observe their consequences.

For example,
an equivalence relation $=_s$, called
\emph{push-equivalence}%
\footnote{%
    This name comes from \cite{paolo_confluence_2009}.
  }
(\cref{figure: pull}),
has been proposed
\cite{vaux_-calcul_2007,paolo_confluence_2009}.
Rules like this were
devised in the context of explicit substitution calculus
\cite{di_cosmo_strong_1997,di_cosmo_strong_1999} and were also used in the
proofs of strong normalization that do not depend on confluence
\cite{accattoli_linear_2013}. 

Consider two $\lambda$-terms, $y\kern1pt x$ and $z$, typed as follows
(an example due to L. Vaux),
%
\begin{align*}
y\,\mathord{:}\, A \to B, z\,\mathord{:}\,A, x\,\mathord{:}\, A &\>\vdash\>
y\kern1pt x\,\mathord{:}\, B\\
y\,\mathord{:}\, A\to B, z\,\mathord{:}\,A &\>\vdash\> z\,\mathord{:}\, A
\end{align*}
and the following two proof nets:

\begin{enumerate}
\item[(a)] the net corresponding to an explicit substitution form of $(y\kern1pt x)[x:=z]$
(\cref{figure: push_ex}(a-1)), and
\item[(b)] the net corresponding to a $y\kern1pt z$ ($\equiv (y\kern1pt x)[x:=z]$ as
  implicit substitution) (\cref{figure: push_ex}(b-1)).
\end{enumerate}
%
Both nets should lead to the same normal form after cut elimination.
However, 
the resulting proof nets (\cref{figure: push_ex}(a-2) and
\cref{figure: push_ex}(b-2)) are not identical.
Push-equivalence is introduced to handle such discrepancies and make
them equivalent.
In the example shown, one can transform one structure into the other via 
$?w$-push-equivalence (\cref{figure: pull}(b))
and $?c$-$?w$-fusion (\cref{figure: cont_weak}).

\begin{figure}
  \begin{center}
  \begin{tikzpicture}
    \node[anchor=center] (labela1) at (-3,1.8) {(a-1)};
    \node[anchor=center] (labelb1) at (4,1.8) {(b-1)};

    \node[anchor=center] (a) at (0,0) {\scalebox{0.28}{\input{figures/5/1-1.tex}}};
    \node[anchor=center] (b) at (6,0) {\scalebox{0.28}{\input{figures/5/2-1.tex}}};

    \node[anchor=center] (am) at (0,-1.8) {\Large$\downarrow^*$};
    \node[anchor=center] (bm) at (6,-1.8) {\Large$\downarrow^*$};

    \node[anchor=center] (labela2) at (-3,-2.5) {(a-2)};
    \node[anchor=center] (labelb2) at (4,-2.5) {(b-2)};

    \node[anchor=center] (a2) at (0,-4.5) {\scalebox{0.28}{\input{figures/5/1-2.tex}}};
    \node[anchor=center] (b2) at (6,-4.2) {\scalebox{0.28}{\input{figures/5/2-2.tex}}};
  \end{tikzpicture}
\end{center}
  \caption{Two nets \ueda{((a-1) and (b-1)) whose normal forms ((a-2) and (b-2))}
need push-equivalence to regard as equivalent.}
  \label{figure: push_ex}
\end{figure}

  However, they come with some subtleties, and the precise definition
  (e.g., whether it is an equivalence relation or should be applied
  only in one way) varies in the literature.
In this section, we introduce these rules 
into our encoding
and verify 
how they affect
confluence and normalization when applied in parallel with 
other cut elimination rules.

\begin{figure}[t]
    \newlength{\pullheight}
    \settoheight{\pullheight}{\scalebox{0.5}{\input{figures/5/cont_pull.tex}}}

    \begin{minipage}{0.495\hsize}
    \vspace{0.5em}
    \begin{center}
        \scalebox{0.54}{\input{figures/5/cont_pull.tex}}
           \raisebox{0.7\pullheight}{$=_s$}
        \scalebox{0.54}{\input{figures/5/cont_pull2.tex}}\\[-9pt]
    (a) $?c$-push-equivalence
    \end{center}
    \end{minipage}
\hfill
    \begin{minipage}{0.495\hsize}
    \vspace{0.5em}
    \begin{center}
        \scalebox{0.54}{\input{figures/5/weakn_pull.tex}}
           \raisebox{0.7\pullheight}{$=_s$}
        \scalebox{0.54}{\input{figures/5/weakn_pull2.tex}}\\[-9pt]
    (b) $?w$-push-equivalence
    \end{center}
    \end{minipage}
  \caption{Push-equivalence.}
  \label{figure: pull}
\end{figure}

\begin{figure}[t]
  \newlength{\cwheight}
  \settoheight{\cwheight}{\scalebox{0.5}{\input{figures/5/cont_weakn.tex}}}

  \centering
  \scalebox{0.54}{\input{figures/5/cont_weakn.tex}}
    \raisebox{0.4\cwheight}{$\to$}
  \quad
  \scalebox{0.54}{\input{figures/5/cont_weakn2.tex}}
  \caption{$?c$-$?w$-fusion.}
  \label{figure: cont_weak}
\end{figure}

\begin{figure}[t]
  \centering
%
  \begin{lstlisting}
  contraction_pull@@  // ?c_{pull}
  {'!'(X1,X2), '?c'(I,X5), {+I,+X3,+X4}, $p[X1,X3,X4|*X]}
    :- {'!'(X1,X2), $p[X1,X3,X4|*X]}, '?c'(I,X5), {+I,+X3,+X4}.

  contraction_push@@  // ?c_{push}
  {'!'(X1,X2), $p[X1,X3,X4|*X]}, '?c'(I,X5), {+I,+X3,+X4},
    :- {'!'(X1,X2), $p[X1,X3,X4|*X], '?c'(I,X5), {+I,+X3,+X4}}.

  weakening_pull@@    // ?w_{pull}
  {'!'(X1,X2), '?w'(X3), $p[X1|*X]} 
    :- {'!'(X1,X2), $p[X1|*X]}, '?w'(X3).

  weakening_push@@    // ?w_{push}
  {'!'(X1,X2), $p[X1|*X]}, '?w'(X3) 
    :- {'!'(X1,X2), '?w'(X3), $p[X1|*X]}.

  // helper
  contraction_weakening_fusion@@
  '?c'({+C1, +C2}, C3),'?w'(C2) :- C1=C3.
\end{lstlisting}
  \vspace*{-0.5em}
  \caption{LMNtal encoding of each rule in \cref{figure: pull} 
and \cref{figure: cont_weak}.}
  \label{code: pull}
\end{figure}

Figure~\ref{code: pull} shows the LMNtal encoding of \cref{figure: pull} 
and \cref{figure: cont_weak}.
We represented the equivalence relation as 
a pair of symmetric
rewrite rules.
The left-to-right direction (moving elements out of the promotion
box) is often called \textit{pull}, and the right-to-left direction (putting
elements into the promotion box) is often called \textit{push}.
We represent them as a pair of one-step rewrite rules rather than as
structural congruence 
because we 
want to analyze each direction of rewriting separately.
We also note that, as an important property of LMNtal,
  a rewrite rule with its LHS and RHS flipped is a syntactically
  correct rewrite rule (unlike \ueda{many}
  other rewrite systems whose LHSs must
  usually be atomic).
%

\begin{figure}[t]
    \begin{center}
        \scalebox{0.34}{\input{figures/5/lam2.tex}}
    \end{center}
    \vspace{-1em}
  \caption{Proof net corresponding to $(\lambda f \,\mathord{:} \, n\mathord{\to} n \: . \: \lambda x \,\mathord{:}\, n \: . \: f \:(f\: x) )\:(\lambda x \,\mathord{:}\, n \: . \: x)$.}
  \label{figure: lam2}
\end{figure}

Using the proof net of \cref{figure: lam2} 
corresponding to $(\lambda f \,\mathord{:} \, n\mathord{\to} n \: . \:
\lambda x \,\mathord{:}\, n \: . \: f \,(f\, x))$
$ (\lambda x
\,\mathord{:}\, n \: . \: x)$, 
%
we investigated the changes in the state space (number of states,
transitions, and end states) caused by the addition of each of 
these rules 
to the cut elimination rules. 
\cref{table: pull} shows the results.
Note that the transitions by the helper rule $?c$-$?w$-fusion 
(as with the transitions during API calls)
have been
abstracted away when constructing the state space using
LaViT.

\begin{table}
	\centering
  \caption{Number of states when each rule of 
  \cref{code: pull} is included/excluded.}
  \label{table: pull}
  \vspace*{-0.5em}
  \begin{center}
     \begin{tabular}{c|c|c|c|c|c|c|c|c} \hline
       \multirow{2}{*}{\#}&
       \multirow{2}{*}{$?c_{\text{pull}}$}&
       \multirow{2}{*}{$?c_{\text{push}}$}&
       \multirow{2}{*}{$?w_{\text{pull}}$}&\multirow{2}{*}{$?w_{\text{push}}$} &
        \# of & \# of tran-& \# of end & \multirow{2}{*}{\takyurevise{SN}} \\
       &&&&& states & sitions & states &  \\ 
\Xhline{1pt}
       1 &              &              &  &  &  452 & 1536 & 1 & \takyurevise{$\checkmark$} \\ \hline
       2 & $\checkmark$ &              &  &  & 1268 & 5246 & 1 & \takyurevise{$\checkmark$} \\ \hline
       3 &              & $\checkmark$ &  &  &  452 & 1536 & 1 & \takyurevise{$\checkmark$} \\ \hline
       4 & $\checkmark$ & $\checkmark$ &  &  & 1268 & 5906 & 1 & \takyurevise{$\checkmark^{\ast}$} \\ \hline
       5 &  &  & $\checkmark$ &              & 592      &   2264 & 1 & \takyurevise{$\checkmark$} \\ \hline
       6 &  &  &              & $\checkmark$ & 25118    & 138827 & 2 & \takyurevise{$\checkmark$} \\ \hline
       7 &  &  & $\checkmark$ & $\checkmark$ & $\infty$ &    --- & --- &  \\ \hline
       \noalign{\vskip 3pt}
       \multicolumn{9}{r@{}}{%
    \takyurevise{\small $^{\ast}$Strong normalization 
    \ueda{up to pull/push equivalence; see text.}}}
    \end{tabular}
  \end{center}
\end{table}

Firstly, we added the $?c$-push-equivalence rules (Rows 2--4).
For $?c$, confluence and strong normalization are proved in
\cite{di_cosmo_strong_1999,accattoli_linear_2013}. 
Using LaViT, 
it is easy to see that these properties are preserved.
When each 
rule
was added one by one (Rows 2 and 3), they all reached a single
end state in a finite number of steps.
When both pull and push
rules were added
(Row 4),
\takyurevise{the push rule introduces additional symmetric
\ueda{(back-and-forth)}
transitions.
Since pull and push are mutually inverse, we regard the pull/push
pair as an equivalence relation and assess strong normalization
\ueda{up to} this equivalence.
Under this view, strong normalization
\ueda{was}
preserved, and the number
of states was preserved from Row 2.}


For $?w$, things are more complicated.
Indeed,
some previous work 
did not 
consider the moving of $?w$
(e.g., \cite{di_cosmo_strong_1999}),
and some allowed pulling only
(e.g., \cite{accattoli_linear_2013}). 
For pull, it was confirmed that both confluence and strong
normalization were maintained, meaning that 
it is a rule that can be safely handled (Row 5). 
This result is consistent with the results by \cite{accattoli_linear_2013}.
However, when push was added, the number of states increased,
and confluence was lost \takyurevise{while strong normalization was
still preserved} (Row 6).
When both pull and push were added,
the state space exploded
(Row 7).
\takyurevise{%
\ueda{This means that, unlike
the $?c$ case (Row 4), strong
normalization 
was genuinely lost, that is, the divergence 
could not 
be absorbed
into the pull/push equivalence.}}
%
\takyurevise{
This result is consistent with Guerrini's counterexample 
\cite{accattoli_kesner_2012} (p.~21),
\ueda{formulated in the setting of $\lambda$-terms with explicit 
substitutions,}
\ueda{where the
structural equivalence
$\equiv_n$ including both $?w$-push and $?w$-pull
caused the loss of strong normalization.}}
%
%
Note that 
$?w_\textrm{push}$
is an
interaction between
disconnected graph elements
(a $?w$-cell and a box), 
which may have properties different from the rewriting of connected
components.

For example, 
suppose
a box and a $?w$ cell are connected 
by a $cut$ cell, as shown in \cref{figure: cause}(a).
If we apply
$?w$-push, 
the $?w$ cell is moved into the box as shown in 
\cref{figure: cause}(b).
In the proof tree, this operation corresponds to (wrongly) moving the $?w$ cell
horizontally 
from a different sub-sequent connected by a cut.

A sufficient condition to ensure the correctness of
$?w$-push-equivalence is to ensure that the target $?w$ cell belongs to the
same sequent as the other auxiliary ports of the target box.

\begin{figure}
  \begin{center}
  \begin{tikzpicture}
    \node[anchor=center] (a) at (0,0) {\scalebox{0.55}{\input{figures/5/cause1.tex}}};
    \node[anchor=center] (labela) at (0,-2) {(a)};

    \node[anchor=center] (seqa) at (4.5,0) {$
          \begin{prooftree}[rule margin=0.4ex]
            \hypo{}
            \infer1[($ax$)]{\vdash A^{\bot}, A}
            \infer1[($?d$)]{\vdash ?A^{\bot}, A}
            \infer1[($!$)]{\vdash ?A^{\bot}, !A}
            \rewrite{\color{red}\box\treebox}
  
            \hypo{ \pi }
            \ellipsis{}{\vdash \Gamma}
            \infer1[($?w$)]{\vdash \Gamma, ?A^{\bot}}
            \rewrite{\color{blue}\box\treebox}
           
            \infer2[($cut$)]{\vdash \Gamma, ?A^{\bot}}
          \end{prooftree}
        $};
      \node[anchor=center] (to) at (8,0) {\Large$\to_{w_{\mathrm{push}}}$};

    \node[anchor=center] (b) at (4,-3) {\scalebox{0.55}{\input{figures/5/cause2.tex}}};
    \node[anchor=center] (labelb) at (4,-5) {(b)};

    \node[anchor=center] (seqb) at (8.5,-3) {$
          \begin{prooftree}[rule margin=0.4ex]
            \hypo{}
            \infer1[($ax$)]{\vdash A^{\bot}, A}
            \infer1[($?d$)]{\vdash ?A^{\bot}, A}
            \infer1[($?w$)]{\vdash ?A^{\bot}, ?A, A}
            \infer1[($!$)]{\vdash ?A^{\bot}, ?A, !A}
            \rewrite{\color{red}\box\treebox}
  
            \hypo{ \pi' }
            \ellipsis{}{\vdash \Gamma'}
            \rewrite{\color{blue}\box\treebox}
           
            \infer2[($cut$)]{\vdash \Gamma, ?A^{\bot}}
          \end{prooftree}
        $};
  \end{tikzpicture}
  \end{center}
  \vspace*{-1em}
  \caption{An incorrect example of $?w$-push.}
  \label{figure: cause}
\end{figure}

To address this issue, we consider introducing a new equivalence 
relation called \textit{$?w$-hook-equivalence} $=_h$ (\cref{figure: hpush}).
This rule is used to associate a box with a $?w$ cell.
When applying $?w$-pull, the $?w$ cell is first hooked (the `\texttt{+}' cell) to the box it
was originally in, and then it is pulled out.
$?w$-push can only be applied by following this hook.
This rule is symmetric and prevents the movement of $?w$ cells
between different sub-sequents by $?w$-push.

\begin{figure}
    \newlength{\pullheightt}
    \settoheight{\pullheightt}{\scalebox{0.5}{\input{figures/5/weakn_pull.tex}}}

    \begin{center}
        \scalebox{0.5}{\input{figures/5/weakn_pull.tex}}
           \raisebox{0.6\pullheightt}{$=_h$}
        \scalebox{0.5}{\input{figures/5/weakn_pull3.tex}}
    \end{center}
    \vspace*{-2em}
    \caption{$?w$-hook-equivalence.}
  \label{figure: hpush}
\end{figure}

\begin{figure}[t]
  \centering
  \begin{lstlisting}
  w_hook_pull@@
  {'!'(X1,X2), '?w'(X3), $p[X1|*X]} :- {'!'(X1,X2), $p[X1|*X], +W}, '?w'(W,X3).
  w_hook_push@@
  {'!'(X1,X2), $p[X1|*X], +W}, '?w'(W,X3) :- {'!'(X1,X2), '?w'(X3), $p[X1|*X]}.

  // helper
  h_ax1@@ +W, '?w'(W,X3) :- '?w'(X3).
  h_ax2@@ {$p, +W}, '?w'(W,X3) :- {$p}, '?w'(X3).
  w_ax@@ '?w'(A,B), '?w'(A) :- '?w'(B).
  c_ax@@ {$p, +A}, {$q, +B}, '?c'({+A,+B},C),'?w'(C,D) :- {$p},{$q}, '?w'(D).

  \end{lstlisting}
  \vspace*{-0.5em}
  \caption{LMNtal encoding of the rule in \cref{figure: hpush}.}
  \label{code: hpush}
\end{figure}

\begin{figure}[t]
  \centering
  \begin{lstlisting}[basicstyle=\ttfamily\small]
  promotion_promotion_hook1@@   // !-!
  {'!'(X1,X2), $p[X1|*X], +W}, cut{+X2,+X3}, {$q[X3,X4|*Y], '!'(X4,X5)}
    :- {{'!'(X1,X2), $p[X1|*X], +W}, cut{+X2,+X3}, $q[X3,X4|*Y], '!'(X4,X5)}.

  promotion_promotion_hook2@@   // !-!
  {'!'(X1,X2), $p[X1|*X]}, cut{+X2,+X3}, {$q[X3,X4|*Y], '!'(X4,X5), +W}
    :- {{'!'(X1,X2), $p[X1|*X]}, cut{+X2,+X3}, $q[X3,X4|*Y], '!'(X4,X5), +W}.

  promotion_dereliction_hook@@  // !-?d
  {'!'(X1,X2), $p[X1|*X], +W}, cut{+X2,+X3}, '?d'(X4,X3)
    :- cut{+X1,+X4}, $p[X1|*X], +W.

  promotion_weakening_hook@@    // !-?w
  {'!'(X1,X2), $p[X1|*X], +W2}, cut{+X2,+X3}, '?w'(X3) 
    :- mell.delete(X1,W), {$p[X1|*X], +W2}, {'?w'(W)}.

  promotion_contraction_hook@@  // !-?c
  {'!'(X1,X2), $p[X1|*X], +W}, cut{+X2,+X3}, '?c'({+C1,+C2}, X3)
    :- mell.copy(X2,A1,A2,A3,B1,B2,C1,C2),{'!'(X1,X2),$p[X1|*X], +W},
       {'?c'({+A1,+A2}, A3)}, {cut{+B1,+B2}}.
  \end{lstlisting}
  \vspace*{-0.5em}
  \caption{LMNtal encoding of cut elimination rules with hooks.}
  \label{code: cut-hook}
\end{figure}

Figure~\ref{code: hpush} shows the LMNtal encoding of \cref{figure: hpush}.
%
\takyurevise{
To ensure that the hook atom `\texttt{+}' does not remain in the end
state, we introduced several helper rules (the last four rules of
\cref{code: hpush}):
\begin{itemize}
  \item \texttt{h\_ax1} and \texttt{h\_ax2} express the equivalence
    between a hooked $?w$ ($\texttt{+W, ?w(W,A)}$) and a plain
    $?w$ ($\texttt{?w(A)}$), for the case where \ueda{the hook
    and the $?w$ cell are at the same level (\texttt{h\_ax1})}
    and the case where
    the hook is inside a membrane (\texttt{h\_ax2}); they are written
    one-way (towards removing $\texttt{+}$) so that no $\texttt{+}$
    remains in the end state.
  \item \texttt{w\_ax} handles the case where the cut-elimination
    rule \mbox{($!\, \mathchar`-\, ?w$)} is applied to a box that
    carries a hook: the hook $\texttt{+}$ is then replaced by a
    fresh $\texttt{?w}$ cell, leaving two $\texttt{?w}$ cells where
    there was originally a single hooked one; \texttt{w\_ax} fuses
    them.
  \item \texttt{c\_ax} handle the case where
    \mbox{($!\, \mathchar`-\, ?c$)} is applied to a box that carries a
    hook: a $\texttt{?c}$ cell then ends up connected to the hook
    $\texttt{+}$, and these rules eliminate that structure.
\end{itemize}}
The transitions due to these helper rules are not essential
and have been
abstracted when constructing the state space using
LaViT.
We also provide rules for detecting cut elimination involving a box
with a hook (\cref{code: cut-hook}).

We investigated whether confluence and strong normalization were
preserved by introducing $?w$-hook-equivalence instead of $?w$-equivalence
to the nets of \cref{figure: lam2}.
The results, shown in \cref{table: hpull}, confirm that
both confluence and strong normalization were
preserved and that the correct end state is reached when
the $?w$-hook-equivalence is added (Row 1)
and when the $?c$-push-equivalence is added simultaneously (Row 2).
%
\ueda{%
The strong normalization of the state space of Row 2 is to be
interpreted as follows.
In the state space of Row 2, there are processes $P_1$ and
$P_2$ such that
(i) $P_1 \longrightarrow P_2$ by ($!\,\mathchar`-\,?c$),
(ii) $P_2 \longrightarrow P_3$ by ($!\,\mathchar`-\,?w$), and
(iii) $P_1 \longrightarrow P_3$ by $?c$-$?w$-fusion,
which forms a 2-cycle in the state space that regards $?c$-$?w$-fusion
as an equivalence.
However, such 2-cycles are artifacts of the abstraction;
$?c$-$?w$-fusion is introduced as a one-way rule that simplifies the net,
and no reduction sequence actually goes back the same concrete net.
Ignoring these artifacts, the result
of Row 2 preserves strong normalization.}

\begin{figure}
  \centering
  \includegraphics[width=1\textwidth]{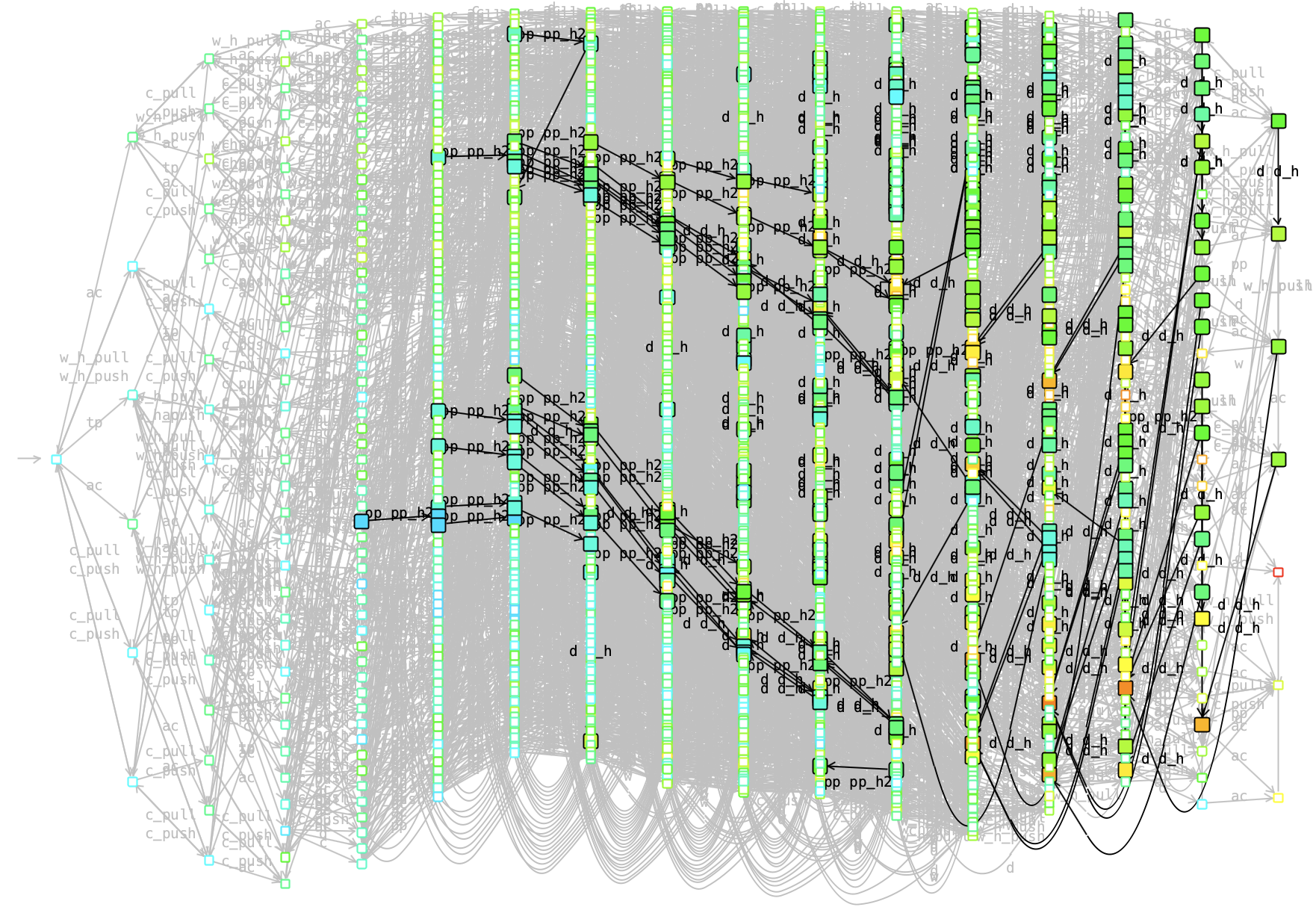}
  \caption{\takyurevise{State space of the example of \cref{figure: lam2} with
$?w$-hook-equivalence and $?c$-push-equivalence.  The dark edges
highlight the transitions whose applied rule is a
cut-elimination on 
\ueda{hooked boxes}
(found via the Rule Name Search
of StateViewer).
Individual 
labels 
\ueda{are shown as shortened rule names but are still}
not intended to be 
\ueda{clearly legible}
at the printed size.}}
  \label{figure: state_hook}
\end{figure}

\takyurevise{%
\ueda{Now note}  
that Row 1 has exactly the same number of states (592) as
Row 5 of \cref{table: pull}.  States in which a box 
carries a
hook are absorbed into their hook-free representatives by the helper
rules.
This equality is as expected: Row 5 of \cref{table: pull} performs
only the safe operations, and the hook mechanism is 
\ueda{to admit only safe $?w$-push,}
so the hooked system
explores the same 
\ueda{state space}
as plain $?w$-pull.
The number of transitions is larger (2440 vs.\ 2264) because cut
elimination on a box carrying a hook is performed by the additional
rules of \cref{code: cut-hook}.}
Figure~\ref{figure: state_hook} highlights the transitions
involving
\ueda{boxes with hooks}
in the state space of Row 2.
%
%
\takyurevise{This is 
\ueda{to qualitatively observe}
how
cut-elimination on hooked boxes is distributed over the 2032-state space;
the correctness of the reached normal form itself is given by the
quantitative results in \cref{table: hpull} (Row 2), not by this
figure.}

\begin{table}
	\centering
  \caption{Number of states when each rule of 
  \cref{code: hpush} is included/excluded.}
  \label{table: hpull}
  \vspace*{-0.5em}
  \begin{center}
    \begin{tabular}{c|c|c|c|c|c|c|c|c} \hline
       \multirow{2}{*}{\#}&
       \multirow{2}{*}{$?c_{\text{pull}}$}&
       \multirow{2}{*}{$?c_{\text{push}}$}&
       \multirow{2}{*}{$?w_{\text{hpull}}$}&\multirow{2}{*}{$?w_{\text{hpush}}$} &
        \# of & \# of tran-& \# of end & \multirow{2}{*}{\takyurevise{SN}} \\
       &&&&& states & sitions & states &  \\ 
\Xhline{1pt}
       1 & &              & $\checkmark$ & $\checkmark$ &  592 & 2440 & 1 & \takyurevise{$\checkmark$} \\ \hline
       2 & $\checkmark$ & $\checkmark$ & $\checkmark$ & $\checkmark$ & 2032 & 11668 & 1 & \takyurevise{$\checkmark^\ast$} \\ \hline
       \noalign{\vskip 3pt}
       \multicolumn{9}{r@{}}{\ueda{$^\ast$ See text.}}
    \end{tabular}
  \end{center}
\end{table}

To sum up,
hook-equivalence 
ensures intended movement of $?w$-cells.
We might be able to find more relaxed sufficient conditions than
hook-equivalence, which is left to future work.
The objective of our workbench is to 
explore such possibilities before formally proving them correct.

\section{Encoding of process calculi}\label{section: encoding-proc}

The proposed language construct for hierarchical graph rewriting,
defined based on the cut elimination of
MELL proof nets, is expected to be 
general enough
to describe systems with box structures
that involve their cloning, migration and deletion.
A typical motivating example is the encoding of the Ambient Calculus
\cite{cardelli_mobile_2000},
a process calculus that deals with box structures called (mobile) ambients
and features replicate (!), \texttt{in},
\texttt{out} and \texttt{open} operations on those box structures.
%
The purpose of this section is to show how we can encode the
copying of mobile ambients where ambient names and their management
play key roles.
%

\begin{figure}[t]
\begin{center}
\resizebox{4.5cm}{!}{\includegraphics{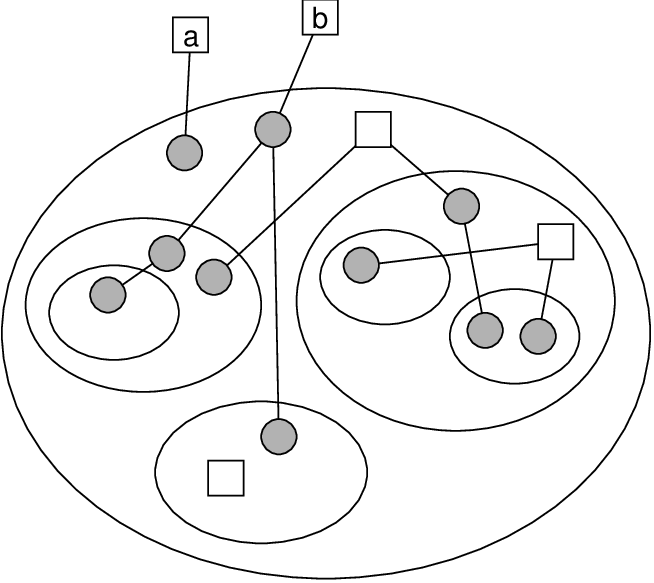}}
\end{center}
\caption{An ambient structure with name trees, where ovals stand for
  ambients, small squares stand for root cells (of which fresh local
  names are abstracted away), 
  and grey circles stands for proxy cells.}
\label{fig:nametree}
\end{figure}

We first briefly review the syntax of the ambient calculus.
Expressions of the ambient calculus are defined as follows, where the
syntactic category $n$ representing names is presupposed: 
\begin{align*}
  \begin{array}{r@{~~}c@{~~}c@{~~}l@{~~}}
     (\textit{processes}) & P & ::= & \zero~~\bigm|~~P \mid P~~\bigm|~~%
(\nu n)P~~\bigm|~~n[P]~~\bigm|~~M.P~~\bigm|~~!P \\[3pt]  
    (\textit{capabilities}) & M & ::= & \texttt{in}\;n ~~\bigm|~~ \texttt{out}\;n ~~\bigm|~~ \texttt{open}\;n \\
  \end{array}
\end{align*}
  Here, $(\nu n)P$ stands for hiding (or the creation of fresh local names);
  $n[P]$ stands for an ambient named $n$;
  $M.P$ stands for a process that performs $M$ and then becomes $P$; and
  $!P$ is a repetition of $P$.
The capabilities are subject to the following reduction rules:
\newcommand{\amb}[2]{#1\mskip1mu[\mskip1.5mu#2\mskip1.5mu]}
\begin{eqnarray*}
\amb{n}{\texttt{in}\ m.P\mid Q} \mid \amb{m}{R}&\>\>\rightarrow\>\>&
   \amb{m}{\amb{n}{P\mid Q}\mid R}\\
\amb{m}{\amb{n}{\texttt{out}\ m.P\mid Q}\mid R} &\rightarrow&
   \amb{n}{P\mid Q}\mid \amb{m}{R}\\
\texttt{open}\ m.P\mid \amb{m}{Q}     &\rightarrow& P\mid Q
\end{eqnarray*}
%
The LMNtal encoding of the ambient calculus is straightforward when it
comes to the nested ambient structure that can be represented
naturally using membranes.  However, 
the representation of ambient names that
can be created fresh
and used for handling
capabilities is far from obvious.  We leave the detailed
discussion to \cite{ueda_encoding_2008-1} but note that each ambient
name is represented as a tree structure, called a \textit{name tree},
comprising \textit{root cells}
and \textit{proxy cells} as depicted in \cref{fig:nametree}.  
A name tree can be thought of representing
what names are referenced
from where (in the distributed and nested ambient structure).
Those name trees are reorganized in an autonomous manner
when the ambient structure is changed by capability operations.
Such explicit representation and management of name references
can be thought of as \textit{refined representation} of the ambient
calculus (in the sense that cut elimination of proof nets models
refined representation of $\beta$-reduction).
%

In our encoding of the ambient calculus
\cite{ueda_encoding_2008-1}, the mapping function $\sema{\cdot}$ for
the syntactic elements other than replication $!P$ is defined as 
in \cref{fig:ambients2lmn},
where
\begin{itemize}
\item $\texttt{@}\textit{amb}$ stands for the encoded ruleset
shown in \cref{code:ambient} which enables local reduction within
individual membranes, and
\item $\downarrow$ is an operator that transforms name
trees into their normal form 
by adjusting the placement of root and proxy cells.
\end{itemize}
\begin{figure}[t]
  \[
  \begin{array}{r@{~~}c@{~~}l}
    \hline\\[-9pt]
    \sema{\zero} & \defeq & \zero \\
    \sema{P\mid Q} & \defeq & (\sema{P} \pc \sema{Q})\downarrow \\
    \sema{(\nu n)P} & \defeq & (\text{hide}_n (\sema{P}\downarrow))\downarrow\, , \;\;\;\text{where}\ \text{hide}_n(\paren{\mem{\texttt{id,name(}n\texttt{),}P}\pc Q}) = \paren{\mem{\texttt{id,}P}\pc Q}\\
    \sema{n[P]} & \defeq & \lmem \texttt{@}\textit{amb} \pc \texttt{amb(}L\texttt{)}\pc \sema{n}(\texttt{L})\pc \sema{P} \rmem \downarrow\\
    \sema{M.P} & \defeq & (\sema{M}(\sema{P}))\downarrow \\ 
    \sema{\textit{op}\;n} & \defeq & \sema{\textit{op}}(\sema{n}) \;\;\;(\textit{op} \in {\texttt{in,out,open}})\\ 
    \sema{\textit{op}} & \defeq & \lambda f . \lambda p. \lpar \textit{op} \paren{\texttt{L,M}}\pc \lmem \texttt{+}M \pc p \rmem \pc f(\texttt{L}) \rpar \;\;\;(\textit{op} \in {\texttt{in,out,open}})\\
    \sema{n} & \defeq & \lambda l. \mem{\texttt{id,name(n),+}l}\\[3pt]\hline
  \end{array}
  \]
\caption{LMNtal representation of ambient expressions.}
\label{fig:ambients2lmn}
\end{figure}
%

For example,
$n[\texttt{open}\ m.P \mid Q] \mid m[R]$ is encoded into
%
\begin{lstlisting}[escapechar=\`]
  {amb(N0), {id, +N0, -N1}, open(M0, {p}), {id, +M0, -M1}, q},
  {amb(M2), {id, +M2, -M3}, r},
  {id, +M1, +M3, name(m)}, {id, +N1, name(n)}.
\end{lstlisting}

LMNtal membranes here play two different
roles: (i) encoding the ambient structure and (ii) managing ambient
names (such as the $n$ above) that are referenced
from various places of the ambient hierarchy and may be freshly created.

Many process calculi have a process copying construct $!P$ 
(defined by the structural congruence rule $!P \equiv P \mid \;!P$). 
The ambient calculus also introduces process copying, but
it is dominantly
used in the form of $!(\texttt{open}\;m.P)$, 
which is described by the following rule:
 \begin{displaymath}
   !(\texttt{open}\;m.P) \mid m[Q] \:\:\to\:\: P \mid Q \mid \;!(\texttt{open}\;m.P)
  \end{displaymath}

One issue that arises in the encoding of $!(\mathrm{open}\; m.P)$ is
that the cloning of $P$ creates new references to the free names
of $P$. Cloning of $\sema{P}$ with free names can be expressed
using aggregates.
This was originally encoded using $\texttt{nlmem.copy}$ but
can now be encoded 
somewhat more concisely 
using $\texttt{mell.copy}$ as follows.
\begin{lstlisting}[escapechar=\`]
  open_repl@@ 
  open_repl(M,{$p}), {amb(M1),{id,+M1,-M2,$mm},$q,@q}, {id,+M,+M2,$m}
    :- mell.copy({$p},A1,A2,A3,B1,B2,remove,P), 
       {cp(A1,A2,A3)}, {B1=B2},
       $q, {id,+M3,$m,$mm}, open_repl(M3,P).

  open_repl_aux@@ remove({$p}) :- $p.
\end{lstlisting}

The operation is done in two steps, the first step 
(\texttt{open\_repl})
for cloning and 
the second step (\texttt{open\_repl\_aux}) for ``opening'' one of the
membranes 
connected to 
\takyurevise{a} 
\texttt{remove} atom
(by rewriting \texttt{\{\$p\}} by \texttt{\$p}) 
.

The complete encoding is shown in \cref{code:ambient}.
The first three rules are one-to-one encoding of \texttt{in},
\texttt{out}, and \texttt{open}, followed by two rules for
replicated \texttt{open} just explained.  The remaining rules are for
the self-adjustment of name trees so that a single proxy
cell is placed in each membrane where the name is referenced directly
or indirectly from its nested membrane.
The final \texttt{gc} rules are for removing unused names.

\begin{figure}
  \begin{lstlisting}[basicstyle=\scriptsize\ttfamily]
  { module(amb).
  
  /* n[in m.P | Q] | m[R] --> m[n[P|Q] | R] */
  in@@
  {amb(N0), {id,+N0,$n}, {id,+M0,-M1,$m0}, in(M0,{$p}), $q,@q},
  {amb(M2), {id,+M2,-M3,$m1}, $r,@r},
  {id,+M1,+M3,$m2} :-
     {amb(M4), {id,+M4,+M5,-M,$m1},
        {amb(N2), {id,+N2,$n}, {id,-M5,$m0}, $p,$q,@q},
     $r,@r},
     {id,+M,$m2}.
  
  /* m[n[out m.P | Q] | R] --> n[P|Q] | m[R] */
  out@@
  {amb(M0), {id,+M0,+M2,$m1}, {id,+N1,$n2},
    {amb(N0), {id,+M1,-M2,$m0}, {id,+N0,-N1,$n}, out(M1,{$p}), $q,@q},
    $r,@r} :-
      {amb(N2), {id,-M3,$m0}, {id,+N2,-N3,$n}, $p,$q,@q},
      {amb(M4), {id,+M3,+M4,$m1}, {id,+N3,$n2}, $r,@r}.
  
  /* open m.P | m[Q] --> P|Q */
  open@@
  open(M,{$p}), {amb(M1), {id,+M1,-M2,$mm}, $q,@q}, {id,+M,+M2,$m} :-
     $p, $q, {id,$m,$mm}.
  
  /* !(open m) | m[Q] --> Q | !(open m) */
  open_repl@@  /* special case of !open */
  open_repl(M,{$p}), {amb(M1), {id,+M1,-M2,$mm},$q,@q}, {id,+M,+M2,$m}
    :- mell.copy({$p},A1,A2,A3,B1,B2,remove,P), {cp(A1,A2,A3)}, {B1=B2},
       $q, {id,+M3,$m,$mm}, open_repl(M3,P).
  
  open_repl_aux@@ remove({$p}) :- $p.
  
  proxy_enter@@
  {$p[M0,M1|*P],@p}, {id,+M0,+M1,$m} :-
     {$p[M0,M1|*P],@p, {id,+M0,+M1,-M}}, {id,+M,$m}.
  
  proxy_resolve@@
  {id,-M,$m0}, {id,+M,$m1} :- {id,$m0,$m1}.
  
  proxy_insert_middle@@
  {{{id,-M,$m},$p,@p},$q,@q} :- {{id,+M0,-M}, {{id,-M0,$m},$p,@p},$q,@q}.
  
  proxy_insert_outer@@
  {{id,+M0,$m0},$p,@p} :- {{id,-M,$m0},$p,@p}, {id,+M0,+M}.
  
  proxy_merge_outer@@
  {id,+M0,$m0}, {id,+M1,$m1}, {{id,-M0,-M1,$m2},$p,@p} :-
       {id,+M,$m0,$m1}, {{id,-M,$m2},$p,@p}.
  
  local_name_in@@
  {$p[M|*P],@p}, {id,+M} :- {{id,+M}, $p[M|*P],@p}.
  
  global_name_out@@
  {{id,name($n),+M0},{$p[M0|*M],@p},$q,@q} :- unary($n) |
      {{id,+M0,-M},{$p[M0|*M],@p},$q,@q}, {id,name($n),+M}.
  
  root_merge@@
  {id,name($n0),$m0}, {id,name($n1), $m1} :-
    unary($n0), unary($n1),
    $n0=$n1 |
    {id,name($n0),$m0,$m1}.
  
  gc1@@ {id} :- .
  gc2@@ {id,name($n)} :- unary($n) | .
  gc3@@ {id,+X,$m}, {{id,-X}, $p,@p} :- {id,$m}, {$p,@p}.
  gc4@@ amb.use :- .
  
  }.
  \end{lstlisting}
  \vspace{-1em}
  \caption{The full encoding of the ambient calculus in LMNtal.}\label{code:ambient}
\end{figure}

\section{Related Work}\label{section: related}

Graph rewriting systems 
enhanced with the notion of hierarchies
have been proposed in various
forms, from mathematical foundations to programming languages 
\cite{berry_chemical_1992,milner_bigraphical_2001,ueda_lmntal_2009,drewes_hierarchical_2002,ene_attributed_2018,alves_new_2011,muroya_hypernet_2020}.
%
The Chemical Abstract Machine (CHAM) by \cite{berry_chemical_1992} and
Bigraphical Reactive Systems (BRS) by \cite{milner_bigraphical_2001} are
graph rewriting formalisms proposed in the context of concurrency theory.
Hierarchical Graph Transformation by \cite{drewes_hierarchical_2002}
introduces hierarchy into the algebraic
graph transformation formalism.
Attributed Hierarchical Port Graphs (AHP) by \cite{ene_attributed_2018}
adds hierarchical representation to the graphical port graph rewriting
tool PORGY by \cite{pinaud_porgy_2012}. 

Hierarchical graph rewriting systems based on proof nets include
Interaction Nets by \cite{lafont_interaction_1989} with boxes
\cite{alves_new_2011} and Hypernet with tokens \cite{muroya_hypernet_2020}. 
These 
proved to be
useful as models of functional languages, but
they are not designed as
general-purpose
modeling languages. 

The correspondence between graph transformation and logical systems is
studied by \cite{torrini_towards_2009} (graph transformation by double pushout 
vs.\ linear logic formulas) and by \cite{ueda_lmntal_2009} 
(conversion of LMNtal expressions into linear logic formulas).
However, the logical interpretation of the box operations is not
yet obvious.

The cloning of boxes in this study atomically clones all graph elements inside
the box.
In the categorical approach to graph transformation, Sesqui-Pushout
(SePO) by \cite{corradini_sesqui-pushout_2006} and PBPO+ by
\cite{overbeek_graph_2021} have been proposed towards graph cloning.
DLGRS by \cite{brenas_verifying_2018} and
BRS by \cite{milner_bigraphical_2001} also provide cloning constructs for
graph elements. 
Other graph rewriting systems that provide copying include Pattern
Graph Rewrite Systems by \cite{kissinger_pattern_2012} derived from the
ZX-calculus by \cite{picturing_quantum_2017}.

In this study, we proposed the use of aggregates of
process contexts to generate an indefinite number of graph
structures for the proper handling of (otherwise dangling) graph edges.
As a 
method of handling an indefinite number of graph elements,
GROOVE by \cite{ghamarian_modelling_2012} and QLMNtal by
\cite{mishina_introducing_2024} 
propose
representations using quantifiers. 
However, the quantifiers in these formalisms are for the
expressive power of an unspecified number of atoms,
whereas the present paper is concerned also with
handling
an unspecified number of free links
in port graph rewriting.


\section{Conclusion and Future Work}

In this work, we extended the syntax of the declarative language
LMNtal based on hierarchical graph rewriting with aggregates of process
contexts, enabled straightforward description of
the operations of promotion
boxes of MELL proof nets, defined operations corresponding to the
cloning and deletion of promotion boxes, and implemented them via 
the 
\texttt{mell} API in the full-fledged implementation.
LMNtal, extended in this way, has become a practical hierarchical
graph rewriting language with 
strong affinity with MELL proof nets. 
Furthermore, by describing several examples, we have demonstrated that
LMNtal serves as a workbench for proof nets 
(with many different formulations)
and that it is possible
to encode concurrency models involving the
reconfiguration
of hierarchical structures. 
Thus, the bidirectional
consideration from proof nets and from hierarchical graph
rewriting languages proved to be beneficial for both sides.

We address future work not mentioned so far.
Firstly,
the cloning of boxes in this study is limited to generating a single clone,
but it might make sense to 
be able to 
generate multiple clones at once.
This
could be realized by extending the \texttt{mell} library,
but
the extension needs non-obvious re-design to allow \texttt{\$a} of
Fig.~\ref{figure: mell} to have an unbounded number of free links.
%
Secondly, we plan to
accommodate hyperlinks with multiple endpoints
\cite{Alim-Access} in our extension.
Since
the structure formed by 
a tree of
$?c$ cells is essentially a
hyperlink, 
it makes sense to allow hyperlinks into our extension,
both theoretically and in practice.
Finally, LMNtal currently does not allow the rewriting of two atoms
connected by a link crossing an indefinite number of membranes 
(``remote reaction'').  Remote reaction across multiple boxes
has been proposed in some 
formulation of proof nets
\cite{accattoli_linear_2013}.
Allowing this requires major language
extension and is a challenging research topic.

\subsubsection*{Acknowledgments}
The authors are indebted to anonymous reviewers for their valuable
comments.
\takyurevise{%
The authors are also grateful to the anonymous reviewers 
\ueda{for their careful reviews}
and to
Lionel Vaux Auclair and Delia Kesner for fruitful discussions, which
helped improve this work.}
This work is partially supported by Grant-In-Aid for Scientific
Research (23K11057, \takyurevise{26K14785}), JSPS, Japan.

\printbibliography

@article{girard_linear_1987,
	title = {Linear logic},
	volume = {50},
	issn = {0304-3975},
	doi = {10.1016/0304-3975(87)90045-4},
	number = {1},
	journal = {Theoretical Computer Science},
	author = {Girard, Jean-Yves},
	year = {1987},
	pages = {1--101},
}

@article{ueda_lmntal_2009,
	title = {{LMNtal} as a hierarchical logic programming language},
	volume = {410},
	issn = {0304-3975},
	doi = {10.1016/j.tcs.2009.07.043},
	number = {46},
	journal = {Theoretical Computer Science},
	author = {Ueda, Kazunori},
	year = {2009},
	pages = {4784--4800},
}

@inproceedings{lafont_interaction_1989,
	series = {{POPL} '90},
	title = {Interaction {Nets}},
	isbn = {0-89791-343-4},
	doi = {10.1145/96709.96718},
	booktitle = {Proceedings of the 17th {ACM} {SIGPLAN}-{SIGACT} {Symposium} on {Principles} of {Programming} {Languages}},
	publisher = {Association for Computing Machinery},
	author = {Lafont, Yves},
	year = {1989},
	pages = {95--108},
}

@incollection{guerrini_proof_2004,
	series = {London {Mathematical} {Society} {Lecture} {Note} {Series}},
	title = {Proof {Nets} and the $\lambda$-{Calculus}},
	booktitle = {Linear {Logic} in {Computer} {Science}},
	publisher = {Cambridge University Press},
	author = {Guerrini, Stefano},
	editor = {Ehrhard, Thomas and Girard, Jean-Yves and Ruet, Paul and Scott, PhilipEditors},
	year = {2004},
	doi = {10.1017/CBO9780511550850.003},
	pages = {65--118},
}

@article{gocho_evolution_2011,
	title = {Evolution of the {LMNtal} runtime to a parallel model checker},
	volume = {28},
	issn = {0289-6540},
	language = {English},
	number = {4},
	journal = {Computer Software},
	author = {Gocho, Masato and Hori, Taisuke and Ueda, Kazunori},
	year = {2011},
	pages = {137--157},
        doi = {10.11309/jssst.28.4_137}
}

@inproceedings{ueda_encoding_2008,
	title = {Encoding the {Pure} {Lambda} {Calculus} into {Hierarchical} {Graph} {Rewriting}},
	isbn = {978-3-540-70590-1},
	booktitle = {Rewriting {Techniques} and {Applications}},
	publisher = {Springer Berlin Heidelberg},
	author = {Ueda, Kazunori},
	editor = {Voronkov, Andrei},
	year = {2008},
	pages = {392--408},
}

@article{ueda_encoding_2008-1,
	title = {Encoding {Distributed} {Process} {Calculi} into {LMNtal}},
	volume = {209},
	issn = {1571-0661},
	doi = {10.1016/j.entcs.2008.04.012},
	journal = {Electronic Notes in Theoretical Computer Science},
	author = {Ueda, Kazunori},
	year = {2008},
	pages = {187--200},
}

@article{_lmntal_2008,
	title = {{LMNtal: The Unifying Programming Language Based on Hierarchical Graph Rewriting}},
	volume = {25},
	doi = {10.11309/jssst.25.1_124},
	number = {1},
	journal = {Computer Software},
	author = {Inui, Atsuyuki and Kudo, Shintaro and Hara, Koji and Mizuno, Ken and Kato, Norio and Ueda, Kazunori},
	year = {2008},
	pages = {1\_124--1\_150},
}

@article{danos_structure_1989,
	title = {The {Structure} of {Multiplicatives}},
	volume = {28},
	doi = {10.1007/bf01622878},
	number = {3},
	journal = {Archive for Mathematical Logic},
	author = {Danos, Vincent and Regnier, Laurent},
	year = {1989},
	pages = {181--203},
}

@inproceedings{ueda_lmntal_2005,
       	author = {Ueda, Kazunori and Kato, Norio},
	title = {{LMNtal}: {A} language model with links and membranes},
        booktitle = {Proc.\ Fifth Int.\ Workshop on Membrane Computing (WMC 2004)},
	series = {LNCS},
	volume = {3365},
	issn = {0302-9743},
	doi = {10.1007/978-3-540-31837-8_6},
	year = {2005},
	pages = {110--125},
}

@article{cardelli_mobile_2000,
	title = {Mobile ambients},
	volume = {240},
	issn = {0304-3975},
	doi = {10.1016/S0304-3975(99)00231-5},
	number = {1},
	journal = {Theoretical Computer Science},
	author = {Cardelli, Luca and Gordon, Andrew D.},
	year = {2000},
	pages = {177--213},
}

@inproceedings{milner_bigraphical_2001,
	booktitle = {12th International Conference on Concurrency Theory ({CONCUR} '01)},
	title = {Bigraphical {Reactive} {Systems}},
	isbn = {3-540-42497-0},
	publisher = {Springer-Verlag},
	author = {Milner, Robin},
	year = {2001},
	pages = {16--35},
    series={LNCS},
    volume={2154},
        doi = {10.1007/3-540-44685-0_2}
}

@article{berry_chemical_1992,
	title = {The chemical abstract machine},
	volume = {96},
	issn = {0304-3975},
	doi = {10.1016/0304-3975(92)90185-I},
	number = {1},
	journal = {Theoretical Computer Science},
	author = {Berry, Gérard and Boudol, Gérard},
	year = {1992},
	pages = {217--248},
}

@article{_lmntal_2010,
	title = {{LMNtal Model Checking using an Integrated Development Environment}},
	volume = {27},
	doi = {10.11309/jssst.27.4_197},
	number = {4},
	journal = {Computer Software},
	author = {Ayano, Takayuki and Hori, Taisuke and Iwasawa, Hiroki and Ogawa, Seiji and Ueda, Kazunori},
	year = {2010},
	pages = {4\_197--4\_214},
}

@article{_lmntal_2005,
	title = {{Design and Implementation of Operation Constructs of Graph Structures in the LMNtal System}},
	volume = {4},
	journal = {Information Technology Letters},
	author = {Kudo, Shintaro and Kato, Norio and Ueda, Kazunori},
	month = aug,
	year = {2005},
	pages = {9--12},
        url = {http://id.nii.ac.jp/1001/00147815/}
}

@article{pagani_strong_2010,
	title = {Strong normalization property for second order linear logic},
	volume = {411},
	issn = {0304-3975},
	doi = {10.1016/j.tcs.2009.07.053},
	number = {2},
	journal = {Theoretical Computer Science},
	author = {Pagani, Michele and Falco, Lorenzo Tortora de},
	year = {2010},
	pages = {410--444},
}

@inproceedings{girard_linear_1993,
	title = {Linear {Logic}: {A} {Survey}},
	author = {Girard, Jean-Yves},
        booktitle="Logic and Algebra of Specification",
        editor = {Bauer, F.L. and Brauer, W. and Schwichtenberg, H.},
	year = 1993,
        series = {NATO ASI Series},
	volume = {94},
        publisher="Springer Berlin Heidelberg",
        pages="63--112",
        isbn="978-3-642-58041-3",
	doi="10.1007/978-3-642-58041-3_3"
}

@article{alves_new_2011,
	title = {A new graphical calculus of proofs},
	volume = {48},
	doi = {10.4204/EPTCS.48.8},
	journal = {Electronic Proceedings in Theoretical Computer Science},
	author = {Alves, Sandra and Fernández, Maribel and Mackie, Ian},
	month = feb,
	year = {2011},
  doi = {10.4204/EPTCS.48.8}
}

@article{drewes_hierarchical_2002,
	title = {Hierarchical {Graph} {Transformation}},
	volume = {64},
	issn = {0022-0000},
	doi = {10.1006/jcss.2001.1790},
	number = {2},
	journal = {Journal of Computer and System Sciences},
	author = {Drewes, Frank and Hoffmann, Berthold and Plump, Detlef},
	year = {2002},
	pages = {249--283},
}

@article{ene_attributed_2018,
	title = {Attributed hierarchical port graphs and applications},
	volume = {265},
	issn = {2075-2180},
	language = {English},
	journal = {Electronic Proceedings in Theoretical Computer Science, EPTCS},
	author = {Ene, Nneka Chinelo and Fernández, Maribel and Pinaud, Bruno},
	month = feb,
	year = {2018},
	pages = {2--19},
        doi = {10.4204/eptcs.265.2}
}

@article{pinaud_porgy_2012,
	series = {Eurographics {Conference} on {Visualization} ({EuroVis} 2012)},
	title = {{PORGY}: {A} {Visual} {Graph} {Rewriting} {Environment} for {Complex} {Systems}},
	volume = {31},
	doi = {10.1111/j.1467-8659.2012.03119.x},
	number = {3},
	journal = {Computer Graphics Forum},
	author = {Pinaud, Bruno and Melançon, Guy and Dubois, Jonathan},
	year = {2012},
	pages = {1265--1274},
}

@inproceedings{takyu_encoding_2023,
	title = {Encoding {MELL} {Cut} {Elimination} into a {Hierarchical} {Graph} {Rewriting} {Language}},
	booktitle = {The 21st {Asian} {Symposium} on {Programming} {Languages} and {Systems} {SRC} \& {Posters}},
	author = {Takyu, Kento and Ueda, Kazunori},
	year = {2023},
}

@inproceedings{mishina_introducing_2024,
	booktitle = {34th {International} {Symposium} on {Logic}-{Based} {Program} {Synthesis} and {Transformation} ({LOPSTR} 2024)},
	title = {Introducing {Quantification} into a {Hierarchical} {Graph} {Rewriting} {Language}},
	author = {Mishina, Haruto and Ueda, Kazunori},
    series={LNCS},
    volume={14919},                  
	year = {2024},
        pages = {220--239},
        doi = {10.1007/978-3-031-71294-4_13}
}

@phdthesis{muroya_hypernet_2020,
	type = {Ph.{D}. thesis},
	title = {Hypernet semantics of programming languages},
	school = {University of Birmingham},
	author = {Muroya, Koko},
	year = {2020},
}

@phdthesis{vaux_-calcul_2007,
	type = {Theses},
	title = {$\lambda$-calcul diff{\'e}rentiel et logique classique : interactions calculatoires},
	url = {https://theses.hal.science/tel-00194149},
	school = {Universit{\'e} de la M{\'e}diterran{\'e}e - Aix-Marseille II},
	author = {Vaux, Lionel},
	month = jan,
	year = {2007},
}

@inproceedings{di_cosmo_strong_1997,
	title = {Strong normalization of explicit substitutions via cut elimination in proof nets},
	doi = {10.1109/LICS.1997.614927},
	booktitle = {Proceedings of {Twelfth} {Annual} {IEEE} {Symposium} on {Logic} in {Computer} {Science}},
	author = {Di Cosmo, R. and Kesner, D.},
	year = {1997},
	pages = {35--46},
}

@inproceedings{di_cosmo_strong_1999,
	title = {Strong {Normalization} of {Proof} {Nets} {Modulo} {Structural} {Congruences}},
	isbn = {978-3-540-48685-5},
	booktitle = {Rewriting {Techniques} and {Applications} (RTA 1999)},
	publisher = {Springer Berlin Heidelberg},
	author = {Di Cosmo, Roberto and Guerrini, Stefano},
    series={LNCS},
    volume={1631},
	year = {1999},
	pages = {75--89},
        doi = {10.1007/3-540-48685-2_6}
}

@article{accattoli_kesner_2012,
	author = {Accattoli, Beniamino and Kesner, Delia},
	title = {Preservation of {Strong} {Normalisation} {Modulo} {Permutations} for the {Structural} {Lambda-Calculus}},
	journal = {Logical Methods in Computer Science},
	volume = {8},
	number = {1},
	year = {2012},
	pages = {1--44},
	doi = {10.2168/LMCS-8(1:28)2012}
}

@inproceedings{accattoli_linear_2013,
	series = {Leibniz {International} {Proceedings} in {Informatics} ({LIPIcs})},
	title = {Linear {Logic} and {Strong} {Normalization}},
	volume = {21},
	isbn = {978-3-939897-53-8},
	doi = {10.4230/LIPIcs.RTA.2013.39},
	booktitle = {24th {International} {Conference} on {Rewriting} {Techniques} and {Applications} ({RTA} 2013)},
	author = {Accattoli, Beniamino},
	editor = {van Raamsdonk, Femke},
	year = {2013},
	pages = {39--54},
}

@inproceedings{overbeek_graph_2021,
	title = {Graph {Rewriting} and {Relabeling} with {PBPO+}},
	isbn = {978-3-030-78946-6},
	booktitle = {Graph {Transformation} (ICGT 2021)},
	publisher = {Springer},
	author = {Overbeek, Roy and Endrullis, Jörg and Rosset, Aloïs},
	editor = {Gadducci, Fabio and Kehrer, Timo},
        series = {LNCS},
	volume = {12741},
	year = {2021},
	pages = {60--80},
        doi = {10.1007/978-3-030-78946-6_4}
}

@inproceedings{brenas_verifying_2018,
	title = {Verifying {Graph} {Transformation} {Systems} with {Description} {Logics}},
	isbn = {978-3-319-92991-0},
	booktitle = {Graph {Transformation}},
	publisher = {Springer},
    series={LNCS},
    volume={10887},
	author = {Brenas, Jon Haël and Echahed, Rachid and Strecker, Martin},
	editor = {Lambers, Leen and Weber, Jens},
	year = {2018},
	pages = {155--170},
        doi = {10.1007/978-3-319-92991-0_10}
}

@article{kissinger_pattern_2012,
	title = {Pattern {Graph} {Rewrite} {Systems}},
	volume = {143},
	doi = {10.4204/EPTCS.143.5},
	journal = {Electronic Proceedings in Theoretical Computer Science},
	author = {Kissinger, Aleks and Merry, Alex and Soloviev, Matvey},
	month = apr,
        pages = {54--66},
	year = {2012},
}

@inproceedings{corradini_sesqui-pushout_2006,
	title = {Sesqui-{Pushout} {Rewriting}},
	isbn = {978-3-540-38872-2},
	booktitle = {Graph {Transformations} (ICGT 2006)},
	author = {Corradini, Andrea and Heindel, Tobias and Hermann, Frank and König, Barbara},
    series={LNCS},
    volume={4178},                  
	year = {2006},
	pages = {30--45},
        doi = {10.1007/11841883_4}
}

@inproceedings{torrini_towards_2009,
	title = {Towards an embedding of {Graph} {Transformation} in {Intuitionistic} {Linear} {Logic}},
	volume = {12},
	doi = {10.4204/EPTCS.12.7},
	booktitle = {Electronic {Proceedings} in {Theoretical} {Computer} {Science}},
	author = {Torrini, Paolo and Heckel, Reiko},
	month = dec,
	year = {2009},
	pages = {99--115},
}

@article{ghamarian_modelling_2012,
        author = "Ghamarian, {Amir Hossein} and {de Mol}, Maarten and Arend Rensink and Eduardo Zambon and Maria Zimakova",
	title = {Modelling and analysis using {GROOVE}},
	volume = {14},
	issn = {1433-2787},
	doi = {10.1007/s10009-011-0186-x},
	number = {1},
	journal = {International Journal on Software Tools for Technology Transfer},
	month = feb,
	year = {2012},
	pages = {15--40},
}

@misc{LMNtal_tutorial,
        author = {Kazunori Ueda},
        title = {Gentle {Introduction} to {LMNtal}: Language {Design} and {Implementation}},
        howpublished = {Tutorial given at the 17th International Conference on Graph Transformation (ICGT 2024), https://conf.researchr.org/details/icgt-2024/icgt-2024-research-papers/17/Gentle-Introduction-to-LMNtal-Language-Design-and-Implementation},
	month = {July},
	year = {2024},
        address = {Enschede, the Netherlands},
        URL = {https://conf.researchr.org/details/icgt-2024/icgt-2024-research-papers/17/Gentle-Introduction-to-LMNtal-Language-Design-and-Implementation}
	}

@article{Alim-Access,
  author={Yasen, Alimujiang and Ueda, Kazunori},
  journal={IEEE Access}, 
  title={Revisiting Graph Types in {HyperLMNtal}: A Modeling Language for Hypergraph Rewriting}, 
  year={2021},
  volume={9},
  number={},
  pages={133449-133460},
  doi={10.1109/ACCESS.2021.3112903}
}

@article{fleury_retore_1994,
  author={Fleury, Arnaud and Retoré, Christian},
  title={The mix rule},
  year={1994},
  volume={4},
  number={2},
  journal={Mathematical Structures in Computer Science},
  pages={273--285},
  doi={10.1017/S0960129500000451},
}

@inproceedings{gonthier_linear_1992,
  author={Gonthier, G. and Abadi, M. and Lévy, J.-J.},
  booktitle={[1992] Proceedings of the Seventh Annual IEEE Symposium on Logic in Computer Science},
  title={Linear logic without boxes},
  year={1992},
  pages={223--234},
  doi={10.1109/LICS.1992.185535}
}

@InProceedings{paolo_confluence_2009,
author="Tranquilli, Paolo",
editor="Gr{\"a}del, Erich
and Kahle, Reinhard",
title="Confluence of Pure Differential Nets with Promotion",
booktitle="Computer Science Logic",
year="2009",
publisher="Springer Berlin Heidelberg",
address="Berlin, Heidelberg",
pages="500--514",
isbn="978-3-642-04027-6",
doi = "10.1007/978-3-642-04027-6_36"
}

@misc{lionel_tutorial,
  author = {Vaux, Lionel},
  title = {Proof nets},
  howpublished = {Tutorial given at the 5th International Workshop on Trends in Linear Logic and Applications (TLLA 2021), https://lipn.univ-paris13.fr/TLLA/2021/},
  month = {June},
  year = {2021},
  address = {Online (Rome virtually)},
  URL={https://lipn.univ-paris13.fr/TLLA/2021/}
}

@book{picturing_quantum_2017, 
  title={Picturing Quantum Processes: A First Course in Quantum Theory and Diagrammatic Reasoning}, 
  publisher={Cambridge University Press}, 
  author={Coecke, Bob and Kissinger, Aleks}, 
  year={2017}
}

@article{takyu_PADL,
	author = {Takyu, Kento and Ueda, Kazunori},
	title = {Enhancing a {Hierarchical} {Graph} {Rewriting} {Language} based on {MELL} {Cut} {Elimination}},
	year = {2025},
      	booktitle = {{Proc.} 27th {International} {Symposium} on {Practical} {Aspects} of {Declarative} {Languages} (PADL 2025)},
    series={LNCS},
    volume={15537},
    pages = {196--214},
    doi = {10.1007/978-3-031-84924-4_13}
}

@InProceedings{sano-icgt2023,
  author="Sano, Jin
  and Ueda, Kazunori",
  editor="Fern{\'a}ndez, Maribel and Poskitt, Christopher M.",
  title={Implementing the $\lambda_{GT}$ Language: A Functional Language with Graphs as First-Class Data},
  booktitle="Proc. 16th International Conference on Graph Transformation (ICGT 2023)",
  series    = {LNCS},
  volume    = {13961},
  year="2023",
  publisher="Springer",
  address="Cham",
  pages="263--277",
  doi="10.1007/978-3-031-36709-0\_14"
}

@article{hyperlmntal,
  author   = {Ueda, Kazunori
and Ogawa, Seiji},
  title    = {{HyperLMNtal}: An Extension of a Hierarchical Graph Rewriting Model},
  journal  = {KI - K{\"u}nstliche Intelligenz},
  year     = {2012},
  day      = {01},
  volume   = {26},
  number   = {1},
  pages    = {27--36},
  doi      = {10.1007/s13218-011-0162-3}
}

@book{EhrigEPT06,
  author    = {Hartmut Ehrig and
               Karsten Ehrig and
               Ulrike Prange and
               Gabriele Taentzer},
  title     = {Fundamentals of Algebraic Graph Transformation},
  series    = {Monographs in Theoretical Computer Science. An {EATCS} Series},
  publisher = {Springer},
  year      = {2006},
  doi       = {10.1007/3-540-31188-2},
  isbn      = {978-3-540-31187-4},
  bibsource = {dblp computer science bibliography, https://dblp.org}
}

@InProceedings{ueda-TACS2001,
  author="Ueda, Kazunori",
  editor="Kobayashi, Naoki and Pierce, Benjamin C.",
  title="Resource-Passing Concurrent Programming",
  booktitle="Proc. 4th International Symposium on Theoretical Aspects of Computer Software (TACS 2001)",
  series    = {LNCS},
  volume    = {2215},
  year="2001",
  publisher="Springer Berlin Heidelberg",
  address="Berlin, Heidelberg",
  pages="95--126",
  isbn="978-3-540-45500-4"
}

@article{CHR,
  author = "Fr{\"u}rwirth, Thom",
  title = "Theory and practice of {Constraint} {Handling} {Rules}",
  journal = "J. Logic Programming",
  volume = "37",
  year = "1998",
  pages = "95--138"
  }

@InProceedings{CHR-LL,
  author="Betz, Hariolf
  and Fr{\"u}hwirth, Thom",
  editor="van Beek, Peter",
  title="A Linear-Logic Semantics for Constraint Handling Rules",
  booktitle="Principles and Practice of Constraint Programming - CP 2005",
  year="2005",
  publisher="Springer Berlin Heidelberg",
  address="Berlin, Heidelberg",
  pages="137--151",
  isbn="978-3-540-32050-0"
}

\end{document}